\documentclass[11pt,%
superscriptaddress,
 amsmath,amssymb,
 aps,
]{revtex4-2}

\renewcommand{\cite}{\citep}
\usepackage[colorlinks=true, linkcolor=blue, citecolor=blue, urlcolor=blue]{hyperref}
\usepackage{setspace}
\usepackage{graphicx}
\usepackage{dcolumn}
\usepackage{bm}

\usepackage{tikz}
\usetikzlibrary{calc}
\usepackage{subcaption}

\usepackage[colorinlistoftodos]{todonotes}

\def\uu{\mathbf{u}}
\def\vv{\mathbf{v}}
\def\ss{\ell}

\usepackage{caption}
\begin{document}
\setstretch{1}
\preprint{APS/123-QED}

\title{Preferential sampling enabled by particle finite size and anisotropic shape}



\author{Helena E. Schreder}
\affiliation{Department of Mechanical Engineering, University of Washington, Seattle, Washington 98195, USA}
\affiliation{Department of Mechanical and Aerospace Engineering, Princeton University, Princeton, New Jersey, 08544}

\author{Kartik Krishna}
\affiliation{Department of Mechanical Engineering, University of Washington, Seattle, Washington 98195, USA}

\author{Steven L. Brunton}
\affiliation{Department of Mechanical Engineering, University of Washington, Seattle, Washington 98195, USA}

\author{Michelle H. DiBenedetto}
\email{mdiben@princeton.edu}
\affiliation{Department of Mechanical and Aerospace Engineering, Princeton University, Princeton, New Jersey, 08544}

\date{\today}

\begin{abstract}
Anisotropic, finite-sized particles, common in environmental and industrial flows, exhibit complex dynamics distinct from those of small, spherical particles. Their shape introduces orientation-dependent forces, and their finite size affects how they experience the flow field.
While the effects of particle inertia are known to cause preferential sampling, in this study we consider whether preferential sampling can arise due to finite size and shape alone by considering inertialess rods using slender-body theory. First, we show analytically that preferential sampling can only occur in this limit given the presence of three ingredients: finite particle size, anisotropic particle shape, and a nonlinear flow field.
Next, to demonstrate this effect, we simulate rods in a steady 2D cellular flow: the Taylor-Green vortex flow. 
By analyzing the rod trajectories, we find that rods do indeed preferentially sample areas of both high and low vorticity, corresponding to the fixed points of the flow: the high vorticity cell centers and the low vorticity saddle points in the corners of each cell. This preferential sampling increases with increasing rod length. We also analyze the same data with respect to the flow's nonlinearity (i.e. curvature) to find that the rods preferentially sample the linear regions of the flow and undersample the nonlinear regions. Finally, we analyze the nonlinear dynamics of this system, showing that chaotic trajectories appear as the rod length increases, and that these chaotic regions in the flow also tend to overlap with higher flow nonlinearity. 
Overall, we show how finite size and anisotropic shape alone can cause particles to preferentially sample a flow field, and that this preferential sampling is highly linked to the flow's nonlinearity. 

\end{abstract}
\maketitle

\section{Introduction}

Environmental flows are complex and varied, and often carry floating material, including logs, debris, and ice.
These floating objects can potentially be tracked using remote sensing~\cite{harms_lagrangian_2024,manucharyan_spinning_2022,lopez-acosta_ice_2019} to aid in measuring environmental flows, where direct observations can be difficult to obtain because of dangerous and unnavigable conditions.
Increased remote sensing of the environment is generating immense quantities of image data of surface processes, including images of floating objects. These objects are forced by the underlying fluid, meaning the trajectories of these objects could potentially be used to reveal flow field information. 
However, they do not necessarily trace the flow as their large size and irregular shape can cause them to deviate from the motion of passive fluid tracers. This non-tracer behavior causes challenges for data interpretation. Also, the objects may cluster or preferentially concentrate, which can further bias the measurements. Thus, we are motivated to study the dynamics of large, non-spherical particles in flows to inform these remote-sensing efforts.

Object transport in environmental flows is related to studies on large particles in particle-laden flows.
For example, one study motivated by floating driftwood in rivers observed that large floating disks and rods will behave like tracers in the potential flow around a cylinder after an initial transient period~\cite{ghaffarian_dynamics_2020}. 
While this suggests that large particles can be reliable tracers, the flow considered in this study was irrotational and largely uniform, and therefore it is not clear the results can be extended to environmental flows which are highly rotational and varied. In contrast, studies which consider large, neutrally-buoyant particles in turbulence have shown that finite-size alters particle dynamics. For example, increasing particle size relative to the Kolmogorov lengthscale tends to reduce particle acceleration variance~\cite{qureshi_turbulent_2007,volk_dynamics_2011}. This can be explained by large particles having a spatial filtering effect which reduces the likelihood of extreme events. In addition, while non-neutrally buoyant particles often exhibit preferential sampling in turbulence~\cite{brandt_particleladen_2022}, neutrally-buoyant, finite-sized, spherical particles have not been observed to cluster~\cite{fiabane_clustering_2012}. This suggests that clustering of spherical particles is largely due to particle inertia and not finite-size.
And yet, many studies of large particles often consider particles which are both large and inertial, and therefore it can be hard to disentangle the two effects~\cite{brandt_particleladen_2022}. It is important to understand these two effects separately, as many floating objects in the environment are not necessarily inertial. For example, large floating ice sheets (floes) can have low inertia relative to the large-scale ocean eddies~\cite{lepparanta_treatise_2012,brenner_scaledependent_2023,kim_characterisation_2025}. In that case, it can be important to understand how finite size alone can alter particle dynamics independent of inertia.

Beyond finite size, irregular shape can also affect particle dynamics. 
Unlike spherical particles, non-spherical particles have orientation-dependent forces and torques. These forces cause even infinitesimally small anisotropic particles to preferentially orient due to velocity gradients~\cite{jeffery_motion_1922,parsa_rotation_2012,pumir_orientation_2011,ni_measurements_2015}.
This added degree of freedom introduces additional complexity. Some studies have developed methods to measure particle rotation as a proxy for information about the flow's vorticity ~\cite{marcus_measurements_2014}.
Other studies have developed methods to relate rigid rod dynamics to velocity differences in order to reconstruct fluid velocity gradient tensors~\cite{cavaiola_assembly_2020} and two-point turbulence statistics~\cite{brizzolara_fiber_2021}.
These examples all rely on the fact that the particles act like tracers and do not preferentially sample the flow; however, as rods increase in size, preferential sampling can potentially become nontrivial.

Preferential sampling is a well-known phenomenon for inertial particles in turbulence  \cite{sudharsan_lagrangian_2016,haller_inertial_2008,sapsis_inertial_2009,eaton_preferential_1994}; heavy particles avoid regions of high vorticity in favor of straining regions~\cite{squires_measurements_1991,brandt_particleladen_2022} while light particles like bubbles in water preferentially concentrate in high-vorticity regions~\cite{mathai_bubbly_2020}. Additionally, inertial small anisotropic particles have been seen to concentrate in low-velocity areas in wall-bounded turbulent flows~\cite{zhang_ellipsoidal_2001,mortensen_dynamics_2008}. 
And yet, there is much less known about how preferential sampling can arise in the absence of particle inertia effects.  
There have been some observations that suggest large, inertialess particles, which are additionally anisotropic, can preferentially sample a flow. In some studies, the particles preferentially sampled high-vorticity regions~\cite{fujiki_attraction_2024,picardo_dynamics_2020,atis_anisotropic_2022}, whereas in another study the particles preferentially sampled low-vorticity regions~\cite{lopez_inertial_2017}.
\citet{lopez_inertial_2017} observed that settling rods oversampled areas of low vorticity in a cellular flow, although the behavior is due largely to their settling dynamics and cannot be attributed solely to finite-size.  
In contrast, neutrally-buoyant deformable particles showed attraction to the center of a vortex in cellular flow~\cite{fujiki_attraction_2024}; similarly, deformable chains of inertialess beads showed attraction to vortices in turbulent flow~\cite{picardo_dynamics_2020}. 
However, both studies observed reduced or no attraction when considering rigid particles. Thus, while anisotropy introduced by particle deformability can enable preferential sampling of inertialess particles, it is still not clear the extent to which rigid, anisotropic particles may preferentially sample.

One study of interest did clearly observe rigid, anisotropic particles preferentially sampling in a shape-dependent way.
\citet{atis_anisotropic_2022} observed experimentally that rods concentrated in a 2D vortical flow. The authors mainly attributed this result to particle inertia. However, the particles in their experiments had extremely low Stokes numbers ($St\le \mathcal{O}(10^{-3})$), and yet only the non-spherical particles were observed to preferentially concentrate. This suggests that non-spherical shape may enable preferential sampling of inertialess particles. 

To isolate the effects of shape and size on the preferential sampling of particles, we consider a simple model of inertialess rods in an ideal flow. The rods are modeled using slender-body theory~\cite{batchelor_slenderbody_1970,cox_motion_1970,olson_motion_1998,shin_rotational_2005}. This model is appropriate for particles with a sufficiently high aspect ratio and has been used to study rods in turbulent flows~\cite{pujara_scaledependent_2019}, other complex flows~\cite{lopez_inertial_2017,subramanian_inertial_2005,cavaiola_assembly_2020}, and quiescent flow~\cite{butler_dynamic_2002,mackaplow_numerical_1998}. 
In contrast to these studies, which focus on orientational dynamics, we instead focus on preferential sampling. We find evidence that inertialess rods preferentially sample a cellular Taylor-Green vortex (TGV) flow.
Studies which use similar systems (i.e., slender-body theory in cellular flow) to analyze particle transport and dispersion have observed that particles have a range of behaviors, including particles trapping in regions of the flow \cite{hu_anomalous_2025,hu_levy_2021,manikantan_subdiffusive_2013}. However, these studies mainly focus on particle transport and dispersion, whereas in this study we present a more thorough investigation of preferential sampling. 
Specifically, we find that particles spend more time on the centers and edges of cells---regions that align with the linear parts of the flow---in a way that leads to preferential sampling. 
We analyze why preferential sampling occurs with respect to the particles' finite size and shape. 

The paper is outlined as follows: in Section~\ref{sec:methods}, we describe the system of interest and derive slender-body theory generically. 
We present our results and discussion in Section~\ref{sec:results} beginning with an analytical discussion of the necessary ingredients for preferential sampling of inertialess particles in any flow (Subsection~\ref{sec:results:3ingredients}). We then present observations of rod behavior in the TGV, including: kinematics (Subsection~\ref{sec:results:kinematics}), preferential alignment (Subsection~\ref{sec:results:prefalign}), preferential sampling areas of vorticity (Subsection~\ref{sec:results:presampvort}), preferential sampling of flow nonlinearity (Subsection~\ref{sec:results:nonlinear}), and chaotic dynamics (Subsection~\ref{sec:results:chaos}). Finally, we end with concluding thoughts in \ref{sec:conclusion}.

\section{Methods}\label{sec:methods}

In this section, we discuss the flow field and particle model which we will use to study preferential sampling of finite size rods. The translational and angular velocities of a rod are derived using slender-body theory. Finally, we decompose the angular velocity into separate components due to fluid vorticity and fluid strain rate.


\subsection{Flow field}\label{sec:methods:flowfield}

\begin{figure}[tb]
    \centering
    \begin{minipage}{0.45\textwidth}
        \centering
        \begin{tikzpicture}
            \node[anchor=south west] (img) at (0,0) {\includegraphics[width=\linewidth]{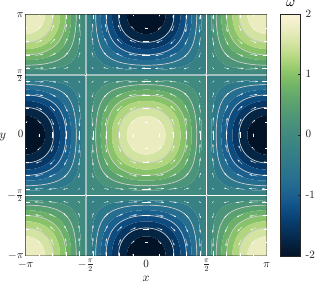}};
            \node[anchor=north west, xshift=10pt, yshift=5pt] at (img.north west) {(a)};
        \end{tikzpicture}
        \label{fig:flowfields}
    \end{minipage}
    \hfill
    \begin{minipage}{0.45\textwidth}
        \centering
        \begin{tikzpicture}
            \node[anchor=south west] (img) at (0,0) {\scalebox{1.2}{\begin{tikzpicture}

    \def\xc{0}  
    \def\yc{0}  
    \def\length{3}  
    \def\perpLength{1.5}  
    \def\soff{0.25} 
    \def\slength{\length/2}
    \def\endlen{0.25} 
    \def\loff{0.3}  

    \def\AngleGamma{110}
    \def\AngleTheta{80}
    \def\AnglePhi{30} 

    \def\fluidvellength{2}
    \def\rodvellength{1}

    \def\xO{-2}
    \def\yO{.7}
    \def\radvort{1}

    \definecolor{colflow}{HTML}{2A5489}
    \definecolor{colrod}{HTML}{ff00ff}
    \definecolor{colannote}{HTML}{515151}
    
    \coordinate (L1) at ({\xc - \length * cos(\AnglePhi)}, {\yc - \length * sin(\AnglePhi)});
    \coordinate (L2) at ({\xc + \length * cos(\AnglePhi)}, {\yc + \length * sin(\AnglePhi)});
    
    \draw[ultra thick,color=black] (L1) -- (L2);


\pgfmathsetmacro{\dxtheta}{\xc - \xO}
\pgfmathsetmacro{\dytheta}{\yc - \yO}

\pgfmathsetmacro{\utx}{-\dytheta}
\pgfmathsetmacro{\uty}{\dxtheta}

\pgfmathsetmacro{\umag}{veclen(\utx,\uty)}
\pgfmathsetmacro{\uscale}{1.2} 
\pgfmathsetmacro{\uunitx}{\uscale * \utx / \umag}
\pgfmathsetmacro{\uunity}{\uscale * \uty / \umag}

\pgfmathsetmacro{\dxtheta}{\xc - \xO}
\pgfmathsetmacro{\dytheta}{\yc - \yO}


\pgfmathsetmacro{\AngleTheta}{atan2(\dxtheta, -\dytheta)} 

\pgfmathsetmacro{\dx}{\xc - \xO}
\pgfmathsetmacro{\dy}{\yc - \yO}

\pgfmathsetmacro{\arcradius}{veclen(\dx,\dy)}

\pgfmathsetmacro{\anglearch}{atan2(\dy,\dx)}

\pgfmathsetmacro{\thetaStart}{\anglearch - 30}
\pgfmathsetmacro{\thetaEnd}{\anglearch + 50}

\draw[->, color=colflow]
  ({\xO + \arcradius*cos(\thetaStart)}, {\yO + \arcradius*sin(\thetaStart)})
  arc[start angle=\thetaStart, end angle=\thetaEnd, radius=\arcradius] node[above] {{\tiny$\text{fluid pathline}$}};

    \def\coordaxlen{.5}
    
    \draw[->,color=colannote] ({\xO}, {\yO}) -- ++(\coordaxlen, 0) node[above] {{\tiny$x$}};
    \draw[->,color=colannote] ({\xO}, {\yO}) -- ++(0, \coordaxlen) node[above] {{\tiny$y$}};


    \fill[color=colflow] (\xO,\yO) circle (2pt);
    \node[anchor=south east,color=colflow] at (\xO, \yO) {{\footnotesize$O$}};

    \coordinate (Uend) at ({\xc + \fluidvellength*cos(\AngleTheta)}, {\yc + \fluidvellength*sin(\AngleTheta)});
    \draw[->,thick,color=colflow] (\xc,\yc) -- (Uend) node[right] {$\mathbf{u}$};
    
    \coordinate (Vend) at ({\xc + \rodvellength*cos(\AngleGamma)}, {\yc + \rodvellength*sin(\AngleGamma)});
    \draw[->,color=colrod,thick] (\xc,\yc) -- (Vend) node[left] {$\mathbf{v}$};

    \draw[thin,color=colannote] ({\xc + \perpLength}, {\yc}) -- (\xc,\yc);
    
\def\radGamma{0.6}
\def\radTheta{0.5}
\def\radPhi{0.7}
\def\arcwidth{thin}

    \draw[<-,\arcwidth,color=colannote] ({\xc + \radGamma*cos(\AngleGamma)}, {\yc + \radGamma*sin(\AngleGamma)}) arc[start angle=\AngleGamma, end angle=\AngleTheta, radius=\radGamma cm];
    \node[color=colannote] at ({\xc -.27+ (\radGamma)*cos((\AngleGamma+\AngleTheta)/2)}, {\yc + (\radGamma)*sin((\AngleGamma+\AngleTheta)/2)}) {\tiny{$\gamma$}};
    
    \draw[<-,\arcwidth,color=colannote] ({\xc + \radTheta*cos(\AnglePhi)}, {\yc + \radTheta*sin(\AnglePhi)}) arc[start angle=\AnglePhi, end angle=\AngleTheta, radius=\radTheta cm];
    \node[color=colannote] at ({\xc + (\radTheta)*cos((\AnglePhi+\AngleTheta)/2)}, {\yc + (\radTheta)*sin((\AnglePhi+\AngleTheta)/2)}) {\tiny{$\theta$}};

    \draw[->,\arcwidth,color=colannote] (\xc+\radPhi,\yc) arc[start angle=0,end angle=\AnglePhi,radius=\radPhi cm];
    \node[color=colannote] at ({\xc + (\radPhi)*cos((\AnglePhi)/2)}, {\yc - .2 + (\radPhi)*sin((\AnglePhi)/2)}) {\tiny{$\phi$}};

    \node[left,xshift=1cm,yshift=-.2cm,color=colannote] at (\xc,\yc) {\tiny{$(x_c, y_c)$}};

    \coordinate (Sstart) at ({\xc - \soff * sin(\AnglePhi)}, {\yc + \soff * cos(\AnglePhi)});
    \coordinate (Send) at ({\xc - \soff * sin(\AnglePhi) - \slength * cos(\AnglePhi)}, {\yc + \soff * cos(\AnglePhi) - \slength * sin(\AnglePhi)});
    \draw[->,color=colannote] (Sstart) -- (Send) node[midway, above] {$\ell$};

    

    
    


    \fill[color=colannote] (\xc,\yc) circle (2pt);
 
\end{tikzpicture}}};
            \node[anchor=north west, xshift=5pt, yshift=-5pt] at (img.north west) {(b)};
        \end{tikzpicture}
        \label{fig:roddiagram}
    \end{minipage}
    \caption{
    (a) TGV flow field where color represents vorticity ($\omega$), arrows represent fluid velocity vectors, and solid lines show streamlines. (b) Schematic of a rod of length $L$, where $(x_c,y_c)$ is its centroid, $\ss$ is the distance along the rod relative to the centroid, $\mathbf{v}$ is the rod velocity, and $\mathbf{u}$ is the local fluid velocity tangent to the fluid pathline. The angles are as follows: $\phi$ is the rod's angle relative to the $x$-axis, $\theta$ is the rod's angle relative to the fluid velocity, and $\gamma$ is the rod velocity's angle relative to the fluid velocity. The signs of  $\theta$ and $\gamma$ are defined relative to vortex center of the cell, denoted by the origin. 
    Angles $\theta$ and $\gamma$ are discussed in more detail in Subsection~\ref{sec:results:prefalign}.
    We also note that (a), as well as all subsequent multi-colored plots, are made using colormaps by~\citet{crameri_scientific_2018}. 
    }
    \label{fig:flowandrod}
\end{figure}
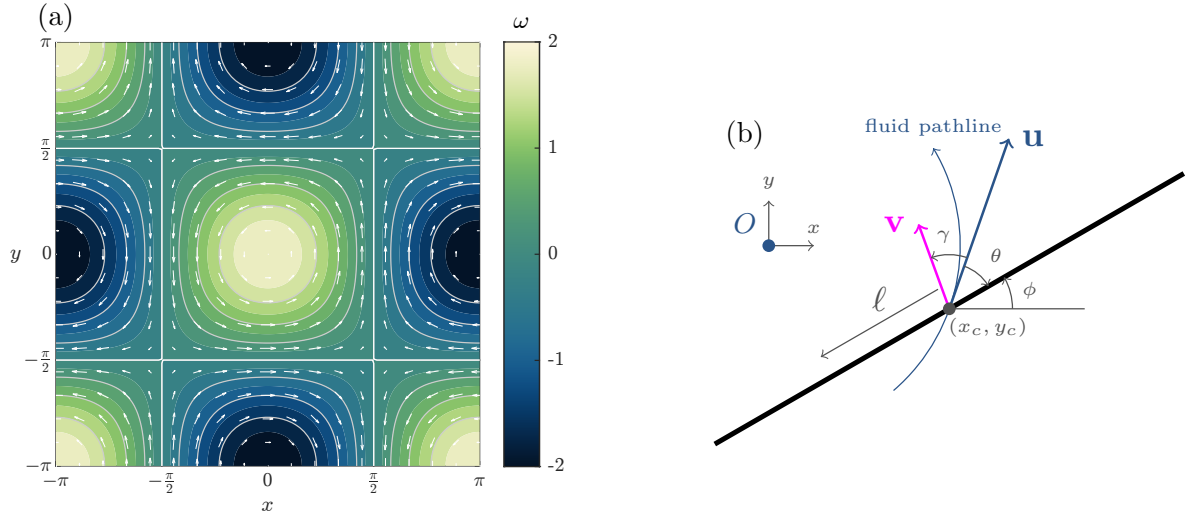

We simulate rods in the two-dimensional, steady Taylor-Green vortex (TGV) flow, also known as cellular flow. The TGV velocity field is shown in Fig.~\ref{fig:flowandrod}(a) and is described by:
\begin{subequations}\label{eq:TGV}
    \begin{align}
    u_x =& -\cos x \sin y \\
    u_y =& \sin x \cos y.
    \end{align}
\end{subequations}
We use the TGV flow because it has a variety of well-suited characteristics for our problem. For example, it includes distinct areas of both high strain and high vorticity, and it is nonlinear spatially. It is also periodic and unbounded which is important because rods often leave their initial cell. Although the TGV is highly idealized, it is often used as a model flow for turbulent eddies. The analytical expression makes for fast simulations so we can run many trials.

\subsection{Rod dynamics}\label{sec:methods:equations}

We derive the equations of motion, including the translational and angular velocity, to simulate the trajectory and orientation for an inertialess rod. Because the rod is rigid, straight, and infinitely thin, we employ slender-body theory~\cite{batchelor_slenderbody_1970} to solve for the forces and torques acting on the rod. While slender-body theory was originally derived for Stokes flow, it has been applied to model the behavior of rods in turbulent flows where the rods are larger than the Kolmogorov length scales~\cite{shin_rotational_2005,pujara_scaledependent_2019}. Even in a turbulent flow, the assumption that rods have negligible inertia remains valid because slender-body theory assumes viscous forces dominate in the region surrounding the rod. This assumption is true given $Re_p\ll1$ where the particle Reynolds number is defined as $Re_p=u_cd/\nu$, and $d$ is the rod's radius, $u_c$ is the characteristic fluid velocity, and $\nu$ is the dynamic viscosity of the fluid. Because we consider rods which are infinitely thin ($d\rightarrow0$), $Re_p\rightarrow0$ and this model will be appropriate no matter the fluid velocity.

Following~\citet{shin_rotational_2005} we calculate the drag force along a rod in the slender-body limit. 
Consider an element of the rod that is a distance of $\ss$ from the rod's centroid (as displayed in Fig.~\ref{fig:flowandrod}(b)); the linear drag force on this element may be defined as:
\begin{equation}\label{eq:forcei}
    f(\ss)=\mathbf{D}(\uu(\ss)-\vv_\ss(\ss)),
\end{equation}
where $\uu(\ss)$ is the local fluid velocity, $\mathbf{D}$ is the drag coefficient tensor which we assume to be constant along the length of the rod, and $\vv_\ss(\ss)$ is the rod's local velocity along its length which varies due to both the rod's angular and translational velocity.
 Alternatively, we can rewrite $\vv_\ss(\ss)$ as a sum of the rod's global velocity components:
\begin{equation}
\vv_\ell(\ss) = \vv + \ss\dot{\mathbf{p}},
\end{equation}
where $\vv$ is now the translational velocity of the rod (i.e. the velocity of the rod's centroid) and $\dot{\mathbf{p}}=\dot{\phi}[-\sin\phi,\cos\phi]$ is the rate of change of the orientation vector of the rod given $\phi$, the angle of the rod measured from the $x$-axis, and $\dot{\phi}$, the angular velocity of the rod. We use this definition to redefine Eq.~\eqref{eq:forcei} in terms of the rod's global velocity components.

We also define the total force along the rod as the sum of the forces on each component of the rod $f(\ss)$ along the total length of the rod $L$. Similarly, we define the total torque acting along the rod. Because the rod is inertialess, the sum of the both the forces and torques must equal zero and we are left with the following systems of equations:
\begin{subequations}\label{eq:forcesbalance}
    \begin{align}
        \sum F &= \mathbf{D}\int_{-L/2}^{L/2}(\uu(\ss)-(\vv +\ss\dot{\mathbf{p}}))\,d\ss = 0,\label{eq:forcesbalance:force}\\
        \sum \tau &= \mathbf{D}\int_{-L/2}^{L/2}\ss\times (\uu(\ss)-(\vv + \ss\dot{\mathbf{p}}))\,d\ss = 0.\label{eq:forcesbalance:torques}
    \end{align}
\end{subequations}
We solve these equations to obtain the rod's translational and angular velocities: 
\begin{subequations}\label{eq:velocity}
    \begin{align}
        \vv &= \frac{1}{L}\int_{-L/2}^{L/2} \uu(\ss) \,d\ss,\label{eq:velocity:translational}\\
         \dot{\phi} &= \frac{12}{L^3}\int_{L/2}^{L/2}\boldsymbol{\ell}\times\mathbf{u}(\ss)\,d\ss.\label{eq:velocity:angular}
    \end{align}
\end{subequations}
Note that $\ss\dot{\mathbf{p}}$ in Eq.~\eqref{eq:forcesbalance:force} and $\boldsymbol{\ell}\times\mathbf{v}$ in Eq.~\eqref{eq:forcesbalance:torques} both integrate to zero over $\ss\in[-L/2,L/2]$ because they are odd functions on a symmetric interval. We derive analytical solutions for each velocity component of a rod in the TGV as a function of centroid location ($x_c,y_c$), rod angle ($\phi$), and rod length ($L$); these equations are reported in Appendix~\ref{ap:anaeqs}.

We can interpret the translational velocity of the rod as the average fluid velocity over the length of the rod, similar to if the rod were spatially filtering the velocity field. The coefficient of the integral for the angular velocity ($12/L^3$) is the inverse of the second moment of area for a rod. 
The angular velocity is discussed further in the next subsection.

\subsection{Angular velocity decomposition}\label{sec:methods:anveldecomp}

\begin{figure}[tb]
    \centering
    \includegraphics[width=\linewidth]{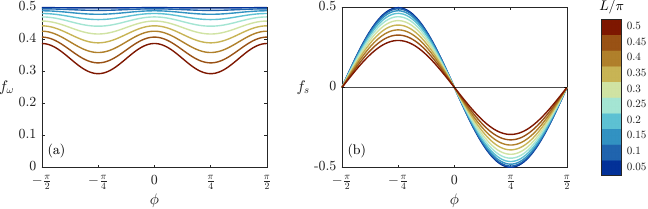}
    \caption{
    Components of the angular velocity ($\dot{\phi} = \omega_c f_\omega(\phi,L) + s_c f_s(\phi,L)$) as functions of rod angle $\phi$ and rod length $L/\pi$.
    (a) Vorticity component $f_\omega$, which is always positive. (b) Strain rate component $f_s$, which is sinusoidal. 
    Both components decrease in magnitude as $L/\pi$ increases.
    }
    \label{fig:angveldecomp}
\end{figure}

We can separate the angular velocity of the rod (as defined in Appendix~\ref{ap:anaeqs}) into distinct contributions from the fluid vorticity and fluid strain:  
\begin{equation}\label{eq:angveldecomp}
    \dot{\phi} = \omega(x_c,y_c) f_\omega(\phi,L) + s(x_c,y_c) f_s(\phi,L),
\end{equation}
where $(x_c,y_c)$ is the rod's centroid position, $\omega$ is the flow vorticity at the centroid, and ${s = 2\sin x\sin y}$ is the normal component of the flow strain rate at the centroid. The functions $f_\omega$ and $f_s$ are independent of the flow and only depend on the rod's orientation and length.  For reference, we  report the functions in Appendix~\ref{ap:anveldecomp}, and we have plotted them in Fig.~\ref{fig:angveldecomp}.
In this plot, we see that the magnitudes of both functions are inversely related to rod length. This indicates, as expected, larger rods will rotate slower than smaller rods at the same location. 

Next, we compare the two functions $f_\omega$ and $f_s$ and their effects on the rod angular dynamics. First, we observe that the vorticity function $f_\omega$ is greater than zero for all rod lengths. This means that the vorticity will always induce rotation on the rod in the same direction as the vorticity given $\omega\ne0$.
In contrast, the strain rate function $f_s$ has fixed points at $\phi=0$ and $\phi=\pm\pi/2$, which correspond to the principal directions of the strain rate in the TGV. The stable fixed point corresponding to the extensional portion of the strain rate is $\phi=0$ for positive $s$ and $\phi=\pm\pi/2$ for negative $s$. In other words, the flow causes the rods to align with the extensional strain rate eigenvectors as expected.

We compare this equation to what is expected as the rods become small by considering the angular velocity of an ellipsoidal, inertialess point particle, which was first described by~\citet{jeffery_motion_1922}. We find that the slender-body formulation recover this equation as we approach the limiting case of infinitesimally small rod. 
In this limit as $L\rightarrow0$, Eq.~\eqref{eq:angveldecomp} can be simplified to:
\begin{equation}\label{eq:angvel_smallL}
    \dot{\phi} = \omega\frac{1}{2} - s\frac{\sin2\phi}{2}.
\end{equation}
This specifically recovers the form of Jeffery's equations for a rod in a 2D incompressible flow field which may be written as $\dot{\phi} = \frac{\omega}{2} - \varepsilon \left( S_{11}\frac{\sin2\phi}{2}+ S_{12}\cos2\phi\right)$ where $S_{11}$ is the normal strain rate, $S_{12}$ is the shear strain rate, and $\varepsilon$ is the particle's eccentricity. In our case, we consider rods with $\varepsilon=1$, and the strain rate in the TGV is $S_{12}=0$ and $S_{11}=s$ meaning we recover Eq.~\eqref{eq:angvel_smallL}.
The correspondence with Jeffery's equations helps to support the validity of the model. 

\subsection{Simulations}\label{sec:methods:simulations}
\begin{figure}[tb]
    \centering
\includegraphics[width=\linewidth]{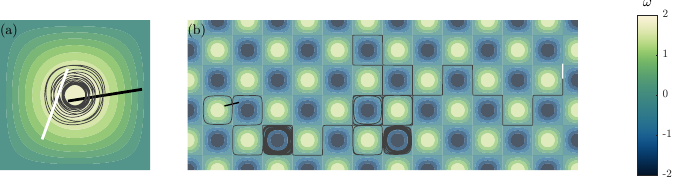}
    \caption{
    Two example trajectories of a rod length $L/\pi=0.5$ in the TGV. 
    The black rod represents the rod's initial position while the white rod represents the final position and the thin grey line traces the trajectory from initial to final. (a) A rod which starts near the center of the cell spirals around the center, moving in and out but never reaching the center. (b) A rod which starts near the edge of the cell leaves its original cell and surfs in between cells, sometimes moving into a cell and spiraling around it multiple times. 
    Animations of these trajectories ((a) \textit{spiraling.mp4} and (b) \textit{surfing.mp4}) are provided in the Supplemental Material~\cite{supplemental}.
    }
    \label{fig:examples}
\end{figure}

We simulated the trajectory and orientation of each one-way coupled inertialess rod in the TGV using Eqs.~\eqref{eq:velocity}. We varied the rod length from $L/\pi =[0.05,0.5]$ where $\pi$ is the length of one TGV cell. This range ensures that the rods are sufficiently large compared to a cell without being so large that their dynamics are too chaotic and their statistics do not converge. Results are composed of statistics from 2100 independent runs for each rod length. Initial rod positions ($x_0$, $y_0$, and $\phi_0$) were chosen randomly, but the same set of initial conditions was used for each rod length. The time integration was performed numerically with a fourth order Runge Kutta integrator and a time step of $\Delta t = 0.01$, where a convergence study was performed to decide the time step. We ran each simulation to $t=1000$, providing enough time for rods to exhibit a variety of behaviors and reach steady statistics. Example rod trajectories are plotted in Fig.~\ref{fig:examples}.

\section{Results and Discussion}\label{sec:results}

In this section, we first detail the necessary conditions for an inertialess particle to preferentially sample a flow. We demonstrate this preferential sampling by studying the behavior of rods in the TGV.
We report their average speed as well as their alignment with the flow.
Next, we examine how the rods preferentially sample regions of flow vorticity and regions of flow curvature. Finally, we examine their behavior using a dynamical systems framework and relate it to their preferential sampling.


\subsection{Three ingredients needed for preferential sampling}\label{sec:results:3ingredients}
To begin, we outline the three basic ingredients needed for preferential sampling in our system. Namely, a system of inertialess particles with linear drag in an incompressible flow field. Although our study and simulations only consider rods, for this discussion we generalize slender-body theory to any 2D particle in a 2D flow by describing the particle translational velocity as the average fluid velocity over the particle's area. This parallels the rod velocity which is the average fluid velocity over the rod's length. We consider only 2D systems here for simplicity, but the results should also apply to 3D.

We find that preferential sampling, or divergent behavior, of the particles in this system can only occur when three ingredients are present, namely: (1) finite particle size, (2) anisotropic particle shape, and (3) a nonlinear flow. We first expand on the importance of (1) finite particle size, which is straightforward: an infinitesimally small, inertialess particle is a tracer particle and thus will never deviate from its initial pathline. Therefore, the particle flow field will remain divergence-free for particles without finite size. 

While finite size allows for our particles to exhibit non-tracer behavior, it does not necessarily mean that particles will be able to preferentially sample the flow. In this system, the particle must have both (1) finite size and (2) anisotropic shape to enable preferential sampling. To demonstrate this, we consider an isotropic, finite-sized particle (i.e. circular in 2D) in an incompressible flow.
We will define its velocity by expanding Eq.~\eqref{eq:velocity:translational} to 2D: $\vv = \frac{1}{A}\int_{S} \uu \,dS$, where $A$ is the object's area and we are evaluating the integral of the velocity over the surface $S=f(x_c,y_c)$. This object acts as a spatial filter of the flow by averaging the fluid velocity over its area. 
If we take the divergence of this velocity, we still find that it will have an incompressible particle flow field given the flow is divergence-free ($\nabla\cdot\uu=0$):
\begin{equation}\label{eq:div_v}
\nabla\cdot\vv = \nabla\cdot\left(\frac{1}{A}\iint_{S} \uu \,dS\right)  = \frac{1}{A}\int_{S}  \nabla\cdot\uu \,dS,
\end{equation}
which gives $\nabla\cdot\vv =0$.
Therefore, even though an isotropic finite-size particle may not exhibit tracer behavior, it can not preferentially concentrate in the flow.
However, Eq.~\eqref{eq:div_v} changes once the particles become anisotropic as the surface integral over the object varies as a function of object orientation (i.e. $S=f(x_c,y_c,\phi)$). This additional variable makes the particle's phase space 3D, enabling particles to cross trajectories in the 2D flow field and exhibit potentially divergent-behavior. 

We have now established that particles which have (1) finite size and (2) anisotropic shape are not constrained to a follow a divergence-free flow field. However, there is still one more ingredient needed to enable preferential sampling, and that ingredient pertains to the flow field. 
Because the particle velocity in this system is a purely linear filter of the flow field, a particle in a linear flow field will behave exactly as if it were a point particle positioned at the centroid of the object, leading to tracer behavior. Therefore, a (3) nonlinear flow is also necessary to allow for preferential sampling of our particles. We demonstrate this by writing the equivalent linearization of a flow about a point of interest, the centroid of the rod $(x_c,y_c)$:
\begin{equation}
\mathbf{u} = \mathbf{u}(x_c,y_c) +(x-x_c)\left.\frac{\partial \mathbf{u}}{\partial x}\right|_{(x_c,y_c)}+(y-y_c)\left.\frac{\partial \mathbf{u}}{\partial y}\right|_{(x_c,y_c)} + \dots
\end{equation}
In the case where $\mathbf{u}$ is a linear flow, the first derivatives are constant and higher order derivatives are zero. This is substituted into Eq.~\eqref{eq:velocity:translational} to obtain:
\begin{equation}
\mathbf{v} = \mathbf{u}(x_c,y_c)+ \frac{\partial \mathbf{u}}{\partial x}\left(\frac{1}{A}\iint_Sx\,dS-x_c\right) + \frac{\partial \mathbf{u}}{\partial y}\left(\frac{1}{A}\iint_Sy\,dS-y_c\right).
\end{equation}
And because $\frac{1}{A}\int_A\mathbf{x}dA$ is the definition of a centroid, all of the higher order terms cancel out and we are left with $\vv = \uu(x_c,y_c)$, or the translational velocity in a linear flow is the fluid velocity at its centroid. This is true for particles of any shape or size---even a finite-sized anisotropic particle needs to be in a nonlinear flow to experience divergent behavior. 

We have demonstrated that divergent behavior in a divergence-free flow (of inertialess particles which follow an expanded slender-body theory) can only occur when particles are both (1) finite size and (2) anisotropic in shape, and are in a (3) nonlinear flow field. 
Finite size is needed so a particle can sample a larger area of the flow than a tracer and can therefore deviate from the pathline it is centered on.
Anisotropic shape gives an extra degree of freedom in the rotation which allows particles to break from a divergence-free flow field.
Finally, nonlinear flow provides the higher order terms necessary for the particle velocity to deviate from the fluid velocity.
The combination of these ingredients allows particles to leave their initial pathlines and sample the flow in a way that tracer particles do not. 
These results are generalizable, and next we will present results for the specific case of rods in the TGV.

\subsection{Rod kinematics}\label{sec:results:kinematics}

\begin{figure}[tb]
    \centering
    \includegraphics[height=2in]{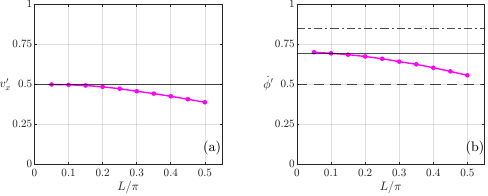}
    \caption{
    RMS (a) translational and (b) angular velocity of rods as a function of length. 
    In both (a) and (b), the solid black line ($-$) shows the RMS velocity for an anisotropic point particle in the TGV after an initial period of transience. 
    In (b), the dashed line ($--$) shows the RMS angular velocity for an isotropic point particle (i.e. rotation solely due to vorticity) and the dashed-dotted line ($-\cdot$) shows the RMS angular velocity if there were no preferential alignment (i.e. an equal distribution of $\phi$). 
    Bootstrapped 95\% confidence intervals for both $u'$ and $\dot{\phi}'$ are $\pm 0.0001$.
    }
    \label{fig:velocity}
\end{figure}

We begin by analyzing rod kinematics as a function of rod length. We find that, on average, larger rods have slower angular and translational speeds than smaller rods. In Fig.~\ref{fig:velocity}, we plot the root mean squared (RMS) $x$-component of the translational velocity $v_x'$ (where the $x$ and $y$ components are symmetrical), along with the angular velocity $\dot{\phi}'$, averaged over all trajectories and trials for each rod length. 
With respect to their translational motion, the rods act like spatial filters of the flow (Eq.~\eqref{eq:velocity:translational}), i.e., as their length increases, they filter over larger areas and average out more extreme fluid velocities. This results in lower average speeds for larger rods, and we show in Fig.~\ref{fig:velocity}(a) that average velocity magnitude ($v_x'$) decreases with increasing rod length ($L/\pi$) as expected. Conversely, as rod length decreases, $v_x'$ approaches the prediction for the RMS velocity of a tracer, also shown in figure ~\ref{fig:velocity}(a).

The rotational statistics of the rods are slightly more complicated than the translational statistics. For example, as rods become smaller, they do not approach the angular velocity of a typical tracer because their anisotropy causes strain-induced rotation. Therefore, we compare the rods' angular velocity in Fig.~\ref{fig:velocity}(b) to the RMS angular velocity of three different types of tracers: an isotropic tracer ($\dot{\phi}'_\text{iso. tracer}$), an anisotropic tracer with random orientation ($\dot{\phi}'_\text{ani. tracer}$), and a simulated anisotropic tracer ($\dot{\phi}'_\text{sim. tracer}$). First, an isotropic tracer particle will only rotate due to the flow's vorticity, so we calculate its angular velocity as $\dot{\phi}'_\text{iso. tracer} = \frac{\omega_\text{flow}'}{2}$ (black dashed line in Fig.~\ref{fig:velocity}(b)). All of the rods in our study rotate faster than this, on average, because the rod anisotropy causes them to rotate in response to strain-induced torques, in addition to vorticity-induced torques.
To quantify rotation due to both vorticity and strain rate, we calculate the RMS value of Eq.~\eqref{eq:angvel_smallL} given a random distribution of $\phi$ (black dashed dotted line in Fig.~\ref{fig:velocity}(b)). 
In this case, $\dot{\phi}'_\text{ani. tracer}$ overpredicts the angular velocity of the simulated rods because the rod's orientation in the simulation is not random. Rather, a rod will preferentially align itself with the flow in a way which will affect its RMS angular velocity.
To account for flow-induced preferential orientation, we run simulations of anisotropic point particles and calculate their RMS angular velocity $\dot{\phi}'_\text{sim. tracer}$  (black solid line in Fig.~\ref{fig:velocity}(b)). We can clearly see that this quantity best matches the limit of small rods. Additionally, we see that this value is lower than the prediction of randomly oriented rods meaning that preferential alignment of the rods in this flow will reduce average particle angular velocities relative to randomly oriented rods. Thus, both increasing rod length and rod alignment with the flow decreases overall angular velocities.

\subsection{Preferential alignment}\label{sec:results:prefalign}

We have just shown how the alignment of rods in the TGV reduces their average angular velocity (Fig.~\ref{fig:velocity}(b)). In this section, we will more closely analyze this preferential alignment behavior and how it relates to the rod kinematics.  

\begin{figure}[tb]
    \centering
    \includegraphics[width=\linewidth]{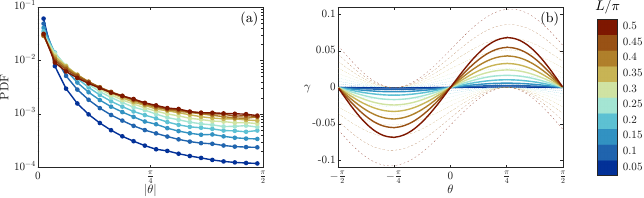}
    \caption{
    (a) Probability density function of $\left|\theta\right|$ as a function of rod length. If $\theta = 0$, the rod is parallel with the flow at its centroid and if it is $\theta = \pm\pi/2$ it is perpendicular. (b) Distribution of $\gamma$ (see Eq.~\eqref{eq:gamma}) for each rod length. Solid lines indicate the mean, $\langle\gamma(\theta)\rangle$, and dotted lines show the full range of $\gamma$ values. As $L\rightarrow0$, $\gamma\rightarrow0$.
    }
    \label{fig:PrefAlignment}
\end{figure}

First, we quantify the degree to which each rod preferentially aligns with the flow. We define $\theta$, shown in Fig.~\ref{fig:flowandrod}(b), as the angle of the rod relative to the fluid velocity at its centroid such that $\theta=0$ when the rod is parallel with the flow and $\theta =\pm \pi/2$ when it is perpendicular to the flow. We plot the probability density functions of the absolute angle $|\theta|$ for each rod length in our simulations in Fig.~\ref{fig:PrefAlignment}(a). From these plots, we observe that all rods tend to have low $|\theta|$ values and are therefore most often oriented parallel to the flow. However, this preferential alignment becomes less extreme as the rods increase in length. This is likely due to the fact that as the rod becomes longer, it interacts with more of the flow field beyond its centroid, further complicating the dynamics.

Next, we compare our results to a related study of rods in a purely rotational flow~\cite{fujiki_attraction_2024}. In this study,~\citet{fujiki_attraction_2024} showed that the direction of a rod's radial motion directly corresponds to the sign of $\theta$. Note that the sign of $\theta$ is determined by comparing the rod's orientation to the flow direction at its centroid: if the rod tilts away from (toward) the vortex center relative to the fluid velocity vector then $\theta>0$ ($\theta<0$). For example, Fig.~\ref{fig:flowandrod}(b) shows a case where $\theta>0$. In a purely rotational flow, if $\theta>0$ then the rod moves away from the center of the vortex and if $\theta<0$ then the rod moves toward the center of the vortex~\cite{fujiki_attraction_2024}.

To compare to the results of \cite{fujiki_attraction_2024}, we need to take into account the fact that the TGV flow is not axially symmetric, and therefore radial velocity does not directly map to the rod's motion away from/toward the center of the cell. Instead, we quantify the rod's relative motion from the center of a cell in the TGV by defining an angle between the fluid pathline and the rod pathline, $\gamma$:
\begin{equation}\label{eq:gamma}
\gamma= \sin^{-1}\left(\frac{|\uu\times\vv|}{|\uu||\vv|}\right).
\end{equation}
For reference, see the diagram in Fig.~\ref{fig:flowandrod}(b). When $\gamma=0$, the rod pathline is following the fluid pathline exactly. When $\gamma>0$, the rod is moving away from the cell center, and when $\gamma<0$ it is moving toward the cell center. Note that $\gamma$ only relates the relative direction of the rod velocity and fluid velocity, not the relative velocity magnitude.

To compare the rod's relative orientation ($\theta$) to its relative motion ($\gamma$) in the cellular flow, we first plot all possible values of $\gamma$ over the TGV as a function of $\theta$ in Fig.~\ref{fig:PrefAlignment}(b). The values are found by calculating $\gamma$ for all rod positions in the cell for each value of $\theta$. The ranges of possible $\gamma$ are plotted as dotted lines, and the averages $\langle\gamma(\theta)\rangle$ are plotted as solid lines.
From this plot, we observe that while the sign of $\langle\gamma(\theta)\rangle$ does correspond to the sign of $\theta$, the sign of any individual value of $\gamma$ is not constrained to correspond to the sign of $\theta$. 
Therefore, unlike in the case of a purely rotational flow considered by~\citet{fujiki_attraction_2024}, the sign of $\theta$ does not always indicate whether the rod is moving toward or away from the center of the vortex. We attribute this deviation to flow nonlinearity, which is present in the TGV but absent in the purely rotational flow. This effect is most pronounced for the largest rods which are more sensitive to the flow's nonlinearity and thus more likely to exhibit sign mismatches.

Another important result from Fig.~\ref{fig:PrefAlignment}(b) is that the magnitude of $\gamma$ increases with rod length. Because the range of possible $\gamma$ for large rods is larger than that for small rods, large rods can deviate from the fluid pathlines more drastically and therefore exhibit more extreme non-tracer behavior. Additionally, we see in this plot that a minimum of $|\gamma|$ is at $\theta = 0$ and that the maxima are at $\theta = \pm\pi/4$. If we recall from Fig.~\ref{fig:PrefAlignment}(a) that large rods are more likely to be misaligned (i.e. $|\theta|>0$), this means that the large rods' reduced preferential alignment will lead to larger $\gamma$. Both of these results together imply that large rods can move across fluid pathlines easier than small rods can. Crossing pathlines is necessary for preferential sampling, and thus this suggests larger rods are more likely to exhibit preferential sampling in this flow.

\subsection{Preferential sampling of high and low vorticity}\label{sec:results:presampvort}

\begin{figure}[tb]
    \centering
    \includegraphics[width=\linewidth]{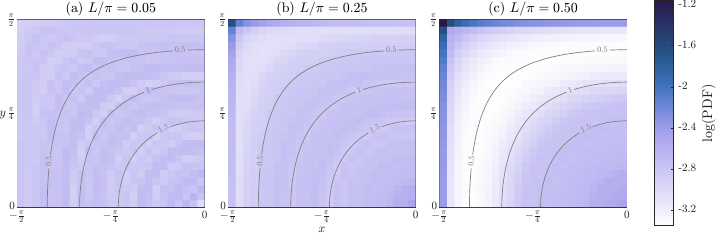}
    \caption{
    Probability density map of rod concentration in the upper-left-corner of the TGV cell, averaged into one cell. Vorticity contours are shown in gray. 
    (a) Smallest rods are evenly distributed across the flow.
    (b) As rod length increases, rods begin concentrating in high-vorticity and low-vorticity regions.
    (c) Largest rod exhibit the strongest preferential sampling, especially in the low-vorticity corner of the cell. 
    }
    \label{fig:conc}
\end{figure}

We next analyze the preferential sampling behavior of the rods with respect to the flow. We first show how the rods preferentially sample the flow by plotting a probability density map of the rods' centroid positions for all trajectories in Fig.~\ref{fig:conc}. Given the symmetry of the flow, we show only the upper-left-most portion of a TGV cell. We see that the smallest rods ($L/\pi=0.05$), which are the most tracer-like, tend to evenly sample the flow as expected. 
In agreement with our necessary ingredients, this even distribution begins to break down as the rods become larger: e.g., an intermediate sized rod, $L/\pi=0.25$ shown in Fig.~\ref{fig:conc}(b), has slight preferential sampling of the edge of the cell which corresponds to low vorticity. This preferential sampling increases for the largest rod we consider, $L/\pi=0.5$ shown in Fig.~\ref{fig:conc}(c), which has strong preferential sampling of the flow. Rods of this size oversample both the center and edges of the cell which correspond to high and low vorticity respectively. Most notably, we see a peak in rod sampling in the corners of the cell. 

\begin{figure}[tb]
    \centering
    \includegraphics[width=\linewidth]{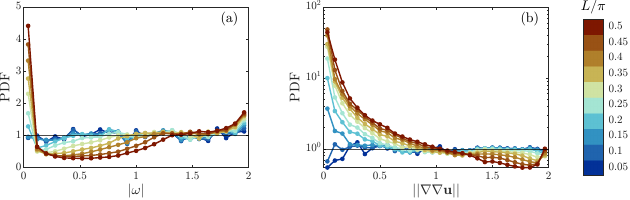}
    \caption{
    Normalized probability density functions (PDF) of rod sampling in the TGV of areas of (a) vorticity ($|\omega|$) and (b) mean curvature ($||\nabla\nabla\mathbf{u}||$) as a function of rod length. A value of one means the rod samples that area at the same rate as a tracer.
    (a) Small rods are evenly dispersed, while larger rods preferentially sample regions of both high and low vorticity.
    (b) Larger rods preferentially sample regions with low curvature (i.e. linear regions).
    }
    \label{fig:hists}
\end{figure}

We can also quantify preferential sampling of the vorticity by the rods. We plot probability distribution functions of vorticity measured at the rods' centroids in Fig.~\ref{fig:hists} as a function of rod length. Again, we clearly see preferential sampling of both high and low vorticity that progressively increases with rod length. Additionally, as rod length increases, the rods increasingly undersample the intermediate vorticity areas.

\begin{figure}[tb]
    \centering
    \includegraphics[width=\linewidth]{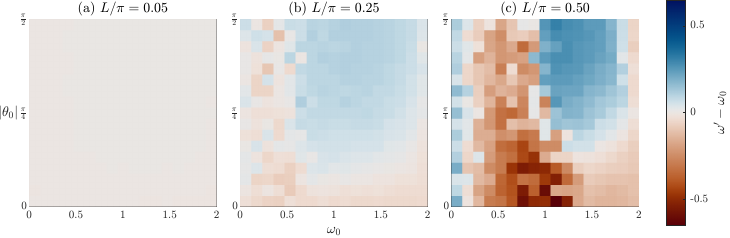}
    \caption{
    Preferential sampling of rods in the TGV as a function of initial angle relative to the streamline, $|\theta_0|$, and initial vorticity, $\omega_0$, where $\omega_0 = 0$ corresponds to the cell edge and $\omega_0 = 2$ to the cell center. Each bin is colored by the average vorticity sampled by rods with those initial conditions, minus their initial vorticity.
    Blue indicates sampling of higher vorticity than the initial value, red indicates sampling lower vorticity, and light colors indicate minimal deviation.
    }
    \label{fig:Phases}
\end{figure}

Along with length, we find that the rod's initial position and initial orientation impact its sampling statistics of the flow. 
Specifically, we observe distinct regions of the initial condition space which cause rods to move to either higher or lower regions of vorticity relative to their initial vorticity. To quantify this behavior, we measure the rod's RMS vorticity relative to its initial position, $\omega'-\omega_0$.
These values are binned by initial vorticity $\omega_0$ and initial relative angle $\theta_0$ and the results are plotted in Fig.~\ref{fig:Phases}. If $\omega'-\omega_0$ is positive, it means the rods on average sampled higher vorticity than their starting point and spent more time near the cell center. If $\omega'-\omega_0$ is negative, it means the rods on average sampled lower vorticity than their starting point and spent more time near the cell edges. The magnitude of $\omega'-\omega_0$ indicates the relative magnitude of the deviation from the rod's initial position. In Fig.~\ref{fig:Phases} we see distinct regions of initial conditions which cause either an increase or decrease in sampled vorticity. For each rod length, if the rod is initially aligned with the flow ($\theta_0 = 0$), it tends to move to regions of lower vorticity (red). If the rod is initially perpendicular to the flow ($\theta_0 = \pi/2$) and starts closer to the center of the cell (high $\omega_0$), it tends to move to regions of higher vorticity (blue). This regime covers much of the initial condition space for $L/\pi = 0.25$ but shrinks for $L/\pi = 0.5$.
Lastly, for larger rods we observe a regime of behavior for low $\omega_0$ (excluding the edge of the cell) where these rods oversample regions of low vorticity. This regime exists for $L/\pi = 0.25$, and increases in size for $L/\pi = 0.5$. 
For all but the smallest rod we see noisier results at low $\omega_0$ which corresponds to the edges of the cell. This indicates these initial conditions could lead to more chaotic trajectories.

These results show that small rods stay close to their initial pathlines, while large rods have more extreme deviations. We also see that large rods are more drawn to the extremes of the flow; specifically, rods which start near the cell's center often move toward the center and rods which start near the cell's edge typically move toward the edge. This causes oversampling of regions of both high and low vorticity and undersampling of regions of intermediate vorticity.
Additionally, because large rods are drawn to the cell edges (with low vorticity), they will more easily move across cells and can be transported farther than small rods.

\subsection{Preferential sampling of nonlinear regions}\label{sec:results:nonlinear}

\begin{figure}[tb]
    \centering
    \includegraphics[width=.5\linewidth]{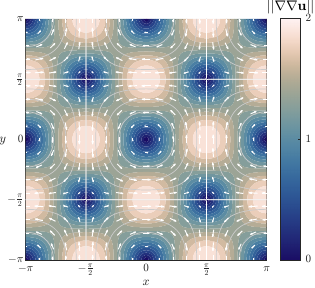}
    \caption{
    Mean curvature ($||\nabla\nabla\mathbf{u}||$) in the TGV represented with a colormap. Quivers represent fluid velocity, and contour lines represent vorticity. Dark blue regions ($||\nabla\nabla\mathbf{u}=0||$) correspond to linear regions of the flow, while light regions  ($||\nabla\nabla\mathbf{u}=2||$) correspond to regions of high nonlinearity.
    }
    \label{fig:CurvMap}
\end{figure}

As we have shown, the length of the rod plays an important role in preferential sampling of high and low vorticity in the TGV. Because this preferential sampling cannot be attributed to particle-inertia effects nor to attracting fixed points in the flow, we revisit our earlier analysis on the necessary three ingredients outlined for preferential sampling in Subsection~\ref{sec:results:3ingredients}. Specifically, we consider the third ingredient: flow nonlinearity. We characterize the nonlinearity of the flow by defining the mean curvature of the flow: 
\begin{equation}
\left|\left|\nabla\nabla\uu\right|\right| = \sqrt{4-s^2-\omega^2},
\end{equation}
where $\left|\left|\nabla\nabla\uu\right|\right|\in[0,2]$ and $\left|\left|\nabla\nabla\uu\right|\right| = 0$ corresponds to a linear region of the flow. A map of the mean curvature of the TGV flow is plotted in Fig.~\ref{fig:CurvMap}. As expected, the linear regions align with the fixed points of the flow: the center fixed points (which are neither attracting nor repelling) at the high vorticity center of each cell and the low vorticity saddle fixed points at the corners of each cell. The most nonlinear areas of the flow exist on the minimum contour of vorticity, corresponding to the edges of the cells in between the saddle points.

To assess the rod preferential sampling with respect to flow nonlinearity, we plot the sampling of the rods' centroids as a function of length. Specifically, we show probability density functions of sampled mean curvature normalized by that of a tracer (Fig.~\ref{fig:hists}(b)). We find that rods show high preferential sampling of the flow's linear regions, and that this preferential sampling behavior increases with rod length. If we compare this rate of preferential sampling of flow curvature to that of flow vorticity (Fig.~\ref{fig:hists}(a)), we clearly see that the mean curvature sampling is orders of magnitude higher than that of vorticity.  This suggests that the preferential sampling of the rods in the flow is more strongly related to the flow's nonlinearity than to its vorticity. 
This may be explained by the fact that non-tracer behavior can only occur in a nonlinear flow, given the model considered in this study. Thus, the preferential sampling may be a result of the rods being less tracer-like in the nonlinear regions, causing them to exhibit more irregular behavior and spend less time in those areas. Instead, rods spend most of their time near the fixed points of the flow, despite neither of these fixed points being attracting. This behavior is described further in Subsection~\ref{sec:results:chaos} 

\subsection{Quasi-periodic and chaotic behavior}\label{sec:results:chaos}

In this section, we study the rod dynamics through the lens of dynamical systems.
Specifically, we investigate how the nature and stability of stagnation points (fixed points) in a flow field dictate the dynamics of a rod with finite size. 
This analysis allows us to potentially generalize our observations to other flow fields.

Because the TGV flow in Eq.~\eqref{eq:TGV} is 2D and incompressible, we can describe it with a streamfunction, resulting in a Hamiltonian system of equations~\cite{aref_frontiers_2017}. While the full equations of motion for the rods are reported in Appendix~\ref{ap:anaeqs}, these apply to the full TGV flow which has an infinite set of fixed points. To make analytical progress, we simplify the periodic TGV down to one cell by Taylor expanding the TGV velocity about $x$ and $y$ and truncating higher order terms that produce more fixed points:
\begin{subequations}
    \begin{align}
        u_x &= -y + \frac{x^2y}{2},\\
        u_y &= x - \frac{xy^2}{2}.
    \end{align}
\end{subequations}
These equations produce a single center fixed point and four saddle points corresponding to the corners of the cell. 

Using Eq.~\eqref{eq:velocity:translational}, we can write the equations for velocity of a rod with finite length in this simplified flow:
\begin{subequations}\label{eq:perturbed}
    \begin{align}
        v_x &= -y_c + \frac{x_c^2y_c}{2}+ \frac{L^2}{24}(x_c\sin2\phi + y_c\cos^2\phi),\\
        v_y &= x_c - \frac{x_cy_c^2}{2}- \frac{L^2}{24}(x_c\sin^2\phi + y_c\sin2\phi).
    \end{align}       
\end{subequations}
We observe that Eqs.~\eqref{eq:perturbed} now take the form of a \emph{perturbed} Hamiltonian  system ($\mathbf{v} = \mathbf{u} + \epsilon g(\mathbf{x}, \phi)$), where $\epsilon = \frac{L^2}{24} \ll1$ agrees with the rod lengths considered in this study; this limit breaks down as the rods grow larger than the cell size. Perturbed Hamiltonian systems can exhibit chaos due to the breaking of heteroclinic connections between saddle fixed points~\cite{ott_chaos_2008}.
This suggests the presence of bifurcations that lead to the onset of a chaotic sea as the perturbation strength is increased; in this case, the perturbation strength is the rod length. This allows for the rods to exhibit chaotic behavior even in a simple, non-chaotic flow. 

As described in Subsection~\ref{sec:results:nonlinear}, our flow field, the TGV, contains two types of fixed points: center fixed points, which are neither attracting nor repelling, located at the center of each cell and saddle fixed points at the corners of each cell.
Numerically, we observe that each type of fixed point produces distinct behaviors in trajectories which are enhanced with rod length. 
Furthermore, we observe several structures in phase space that confirm the perturbed Hamiltonian interpretation of rod dynamics.

\begin{figure}[tb]
    \centering
    \includegraphics[width=\linewidth]{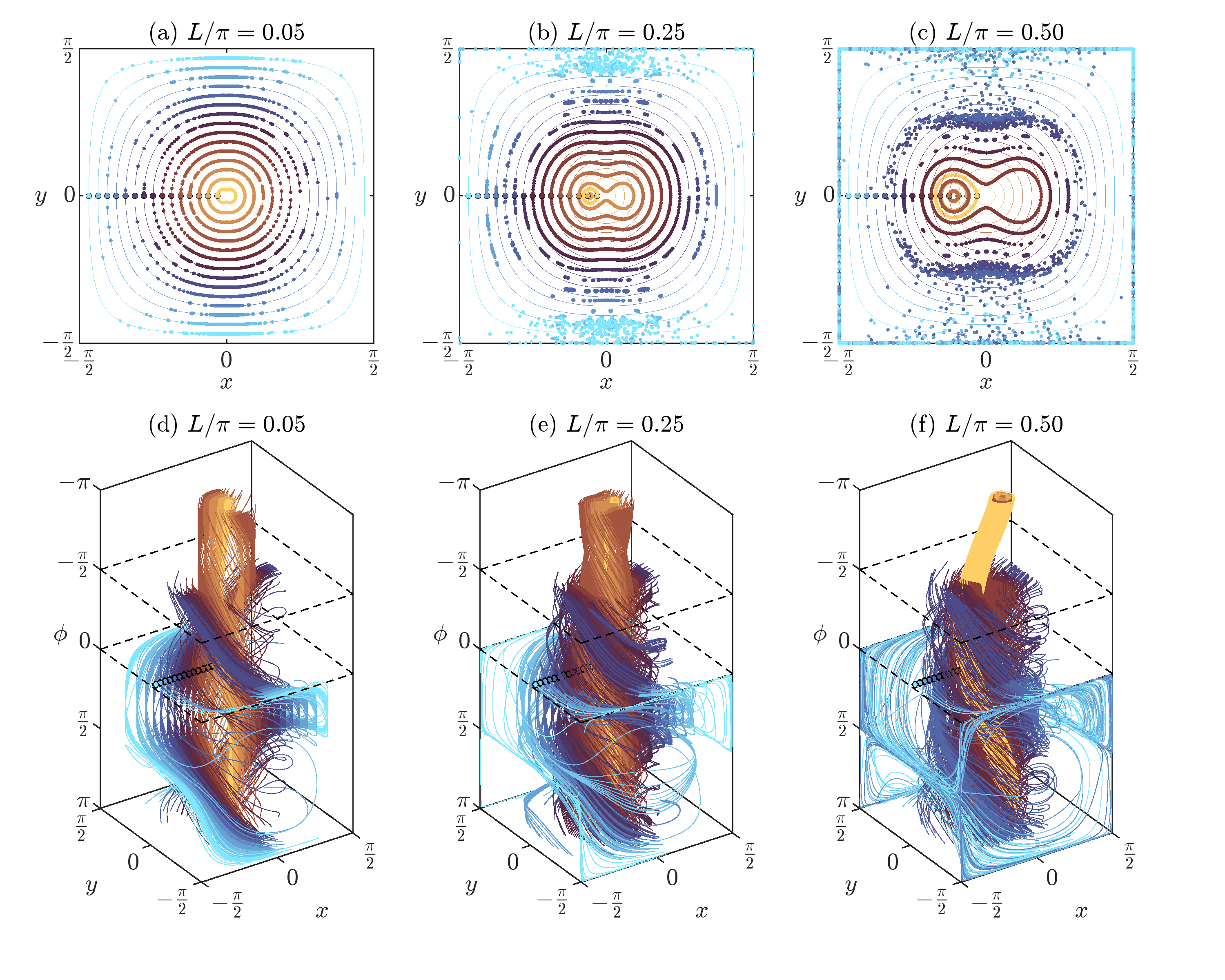}
    \caption{
    (d-f) Trajectories in phase space and (a-c) corresponding Poincaré sections at $\phi=0$ for rods in the TGV with initial conditions $\phi_0=0$, $y_0=0$, and $x_0\in(-\pi/2,0)$. On all plots, initial positions are marked with black outlines, and each color corresponds to a unique trajectory. Rods that left their original cell were reflected into one cell. 
    Thin lines in the background of (a-c) show the fluid streamlines associated with each trajectory, and they are colored to match the trajectories.
    To better visualize internal orbits, some outer trajectories are removed from (d-f); removed sections are indicated with dashed lines. 
    Trajectories that are symmetric about $x=0$ in the Poincaré sections (a-c)---all trajectories in (a), and outer trajectories in (b) and (c)---are approximately $\pi$-periodic in $\phi$ as seen in the 3D phase space trajectories (d-f). Non-symmetric trajectories in (b) and (c) form 3D trajectories in phase space which exhibit $2\pi$ periodicity as seen in (e) and (f).
    }
    \label{fig:Poincare_3D}
\end{figure}

We show evidence of these behaviors in Figs.~\ref{fig:Poincare_3D}(a-c) which show Poincaré sections at $\phi = 0$ for a small number of trajectories, each plotted with a different color. 
The Poincaré sections are accompanied by their full trajectories in phase space plotted in Figs.~\ref{fig:Poincare_3D}(d-f).
We observe that the smallest rods, which act similarly to tracers, appear to have periodic orbits in the $x-y$ plane. However, in phase space, these trajectories are actually quasi-periodic.
As the rod's length increases, the orbits which previously followed the flow streamlines are altered.
Near the cell center, the trajectories begin to focus inward. In contrast, near the cell edges, we observe a breakdown of the orderly trajectories. 
These phase space trajectories near the cell center form distinct lines in the Poincaré sections (Figs.~\ref{fig:Poincare_3D}(a-c)) which imply the presence of Kolmogorov–Arnold–Moser (KAM) tori, meaning that these rods move in a quasi-periodic fashion. These tori are clearly present in the phase space trajectories in Figs.~\ref{fig:Poincare_3D}(d-f). Additionally, we perform a change of variables to plot trajectories on periodic tori in Appendix~\ref{ap:tori}.
Since these trajectories are bound by tori, they will never reach the center nor the edge of the cell. For reference, an example trajectory in real space which has a quasi-periodic orbit is plotted in Fig.~\ref{fig:examples}(a). 

In contrast, we observe that trajectories near the edge of the cells become increasingly chaotic as rod length increases, demonstrated by the chaotic sea of points in Figs.~\ref{fig:Poincare_3D}(b,c).
We also observe that the saddle points cause the rods to move in between cells, ``surfing" from corner to corner (e.g. Fig.~\ref{fig:examples}(b)). 
While these ``surfing" rods may move inward away from the cell edges, they can never cross through a torus.

We can relate this analysis to the preferential sampling discussed earlier. For example, the increased sampling of high-vorticity regions is likely due to rods being bound to KAM tori.
We also see that the regions corresponding to the chaotic trajectories overlap with the nonlinear regions of the flow as shown in the mean curvature map in Fig.~\ref{fig:CurvMap}. Thus, the chaotic regions correspond to areas of lower preferential sampling. This suggests that the transition to chaos may be an indicator of the preferential sampling we observe in this system. 
Finally, it also suggests that phenomena like preferential sampling does not necessarily imply an attracting limit-set, such as a limit cycle for example.

\section{Conclusion}\label{sec:conclusion}
In summary, we have shown that inertialess particles in the slender-body limit can preferentially sample a flow when all of the following three ingredients are met: the particles are finite in size, anisotropic in shape, and in a  flow that is nonlinear.
Under these conditions, particles can cross fluid pathlines and exhibit divergent behavior, even in a divergence-free flow field.

To demonstrate this effect, we simulated rods in the 2D Taylor-Green Vortex flow. We clearly observed that rods preferentially sampled the flow and that preferential sampling increased with rod length. We found that large rods oversampled both high-vorticity regions and low-vorticity regions. 
These preferentially-sampled regions also corresponded to regions of low flow-curvature, i.e. the linear regions of the flow. The sampling bias was far stronger with respect to the flow curvature than to the vorticity, suggesting that flow nonlinearity is the primary mechanism responsible for the observed preferential sampling.
This follows from the requirement that non-tracer behavior may only occur in nonlinear flow.
As rods are more tracer-like in the linear regions, they are more constrained to the fluid pathlines which may cause them to oversample those regions. 

We lastly analyzed the trajectories of rods in phase space and showed that the rods that began near cell centers formed quasi-periodic orbits, while those that started near the cell edges exhibited chaotic trajectories. The chaotic regions aligned with the nonlinear regions of the flow field which also corresponded to the the areas of undersampling.
This implies that the chaos acts to repel the rods towards the oversampled linear regions rather than  acting as an attractor.

The key takeaway from our work is that preferential sampling can arise without particle inertia---anisotropic shape and finite size are sufficient. Many trackable, large objects in environmental flows are both anisotropic and inertial. Disentangling the independent and combined effects of size, shape, and inertia will be necessary to accurately infer flow dynamics from their motion. Unlike inertial preferential sampling, which typically arises from attracting and repelling regions, the behavior we observe occurs without true attractors or repellers. Instead, the difference between the particles' chaotic behavior in the nonlinear regions and the tracer-like behavior in the linear regions produces the observed preferential sampling. While we have isolated finite-size effects in this study, these dynamics are likely modified in the presence of particle inertia, with multiple preferential sampling mechanisms acting simultaneously. Recognizing and quantifying these mechanisms will be critical for interpreting particle and object motion in environmental flows, and for avoiding bias when using anisotropic particles to infer fluid dynamics.

\section{Acknowledgments}\label{sec:acknowledgments}
This material is based upon work supported by the National Science Foundation Graduate Research Fellowship Program under Grant No. DGE-2444107. Any opinions, findings, and conclusions or recommendations expressed in this material are those of the authors and do not necessarily reflect the views of the National Science Foundation. 

\appendix
\section{Analytical solution for velocity of rod in TGV}\label{ap:anaeqs}
The general expressions for the translational and angular velocities of slender-body rods in an arbitrary flow (Eqs.~\eqref{eq:velocity}) are presented in Subsection~\ref{sec:methods:equations}. Here, we derive the specific analytical solutions for a rod of length $L$ in the TGV.

Recall that the rod's centroid is located at $(x_c,y_c)$, its orientation relative to the $x$-axis is $\phi$, and $\ell$ denotes the position along the rod relative to its centroid (see Fig.~\ref{fig:flowandrod}(b) for a schematic drawing). To evaluate the integrals in Eqs.~\eqref{eq:velocity}, we express the spatial coordinates in terms of a particle reference frame:
\begin{equation}\label{eq:particlerefframe}
    \begin{bmatrix} x \\ y
    \end{bmatrix} = 
    \begin{bmatrix} \ell\cos\phi +x_c \\ \ell\sin\phi +y_c  
    \end{bmatrix}.
\end{equation}

We substitute Eqs.~\eqref{eq:particlerefframe} into the TGV velocity field (Eqs.~\eqref{eq:TGV}) to obtain fluid velocity equations which are a function of the position along the rod, $\ell$. These fluid velocities are inserted into Eq.~\eqref{eq:velocity:translational} to find the rod's translational velocity:
\begin{subequations}
    \begin{align}
        v_x &= \frac{1}{L}\int_{-L/2}^{L/2} -\cos(\ell\cos\phi +x_c)\sin(\ell\sin\phi +y_c) \,d\ss,\\
        v_y &= \frac{1}{L}\int_{-L/2}^{L/2} \sin(\ell\cos\phi +x_c)\cos(\ell\sin\phi +y_c) \,d\ss.
    \end{align}
\end{subequations}
Solving these integrals, we obtain solutions for the translational velocity of a rod in the TGV:
\begin{equation}\label{eq:anasol}
    \begin{bmatrix}v_x\\v_y\end{bmatrix} =\begin{bmatrix}\Phi_1(\phi,L) & \Phi_2(\phi,L) \\ \Phi_2(\phi,L) & \Phi_1(\phi,L)\end{bmatrix}\begin{bmatrix}u_x(x_c,y_c)\\u_y(x_c,y_c)\end{bmatrix},
\end{equation}
where we have organized the equations into the components from the fluid velocity at the rod's centroid ($u_x(x_c,y_c)$ and $u_y(x_c,y_c)$) and the rod's orientation and length ($\Phi_1(\phi,L)$ and $\Phi_2(\phi,L)$). $\Phi_1$ and $\Phi_2$ are defined as:
\begin{subequations}
\label{eq:fullroddyn}
\begin{align}
\Phi_1&=\frac{2\sec(2\phi)}{L}\left[\cos\phi\cos\left(\frac{L}{2}\sin\phi\right)\sin\left(\frac{L}{2}\cos\phi\right) - \sin\phi\cos\left(\frac{L}{2}\cos\phi\right)\sin\left(\frac{L}{2}\sin\phi\right)\right], \\
\Phi_2&=\frac{2\sec(2\phi)}{L}\left[\sin\phi\cos\left(\frac{L}{2}\sin\phi\right)\sin\left(\frac{L}{2}\cos\phi\right) - \cos\phi\cos\left(\frac{L}{2}\cos\phi\right)\sin\left(\frac{L}{2}\sin\phi\right)\right].
\end{align}
\end{subequations}

We can similarly solve for the angular velocity of rods in the TGV. We use Eq.~\eqref{eq:velocity:angular}, rewriting $\boldsymbol{\ell}\times\mathbf{u}$ as $\ell(\cos\phi u_y - \sin\phi u_x)$ so that the coordinates along the rod match the fluid velocity, which is expressed in Cartesian coordinates. The integral becomes:
\begin{equation}
    \dot{\phi} = \frac{12}{L^3}\int_{L/2}^{L/2} \ell(\cos\phi \sin(\ell\cos\phi +x_c)\cos(\ell\sin\phi +y_c) + \sin\phi u_x) \,d\ss.
\end{equation}
In solving this integral, we obtain this solution for the angular velocity:
\begin{align}
\dot{\phi} = \frac{12}{L^3}&\left(\cos(x_c-y_c)\left[\frac{\cos\phi+\sin\phi}{\cos2\phi}\sin\left(\frac{L}{2}\left(\cos\phi-\sin\phi\right)\right)-\frac{L}{2}\cos\left(\frac{L}{2}\left(\cos\phi-\sin\phi\right)\right)\right]\right. \notag\\ 
    &+ \left.\cos(x_c+y_c)\left[\frac{\cos\phi-\sin\phi}{\cos2\phi}\sin\left(\frac{L}{2}\left(\cos\phi+\sin\phi\right)\right)-\frac{L}{2}\cos\left(\frac{L}{2}\left(\cos\phi+\sin\phi\right)\right)\right]\right).
\end{align}



\section{Angular velocity decomposition components}\label{ap:anveldecomp}
As discussed in Subsection~\ref{sec:methods:anveldecomp}, the angular velocity of a rod in the TGV can be separated into components due to vorticity and due to strain rate. We plot the functions for each component in Fig.~\ref{fig:angveldecomp}, and here we report the closed forms of these functions:
\begin{subequations}\label{eq:angvelfun}
    \begin{align}
            f_\omega(\phi,L) =\frac{12}{L^3}&\left(\frac{\cos\phi\left[ \sin\left(\frac{L}{2}\cos\phi\right)\cos\left(\frac{L}{2}\sin\phi\right) \right]-\sin\phi\left[ \cos\left(\frac{L}{2}\cos\phi\right)\sin\left(\frac{L}{2}\sin\phi\right) \right]}{\cos2\phi}\right. \notag\\
            &\left.-\frac{L}{2}\cos\left(\frac{L}{2}\cos\phi\right)\cos\left(\frac{L}{2}\sin\phi\right)\right), \\
            f_s(\phi,\ell) =\frac{12}{L^3}&\left(\frac{-\sin\phi\left[ \sin\left(\frac{L}{2}\cos\phi\right)\cos\left(\frac{L}{2}\sin\phi\right) \right]+\cos\phi\left[ \cos\left(\frac{L}{2}\cos\phi\right)\sin\left(\frac{L}{2}\sin\phi\right) \right]}{\cos2\phi} \right. \notag \\
            &\left.- \frac{L}{2}\sin\left(\frac{L}{2}\cos\phi\right)\sin\left(\frac{L}{2}\sin\phi\right)\right) .
    \end{align}
\end{subequations}

\section{Tori}\label{ap:tori}
\begin{figure}[tb]
    \centering
    \includegraphics[width=0.9\linewidth]{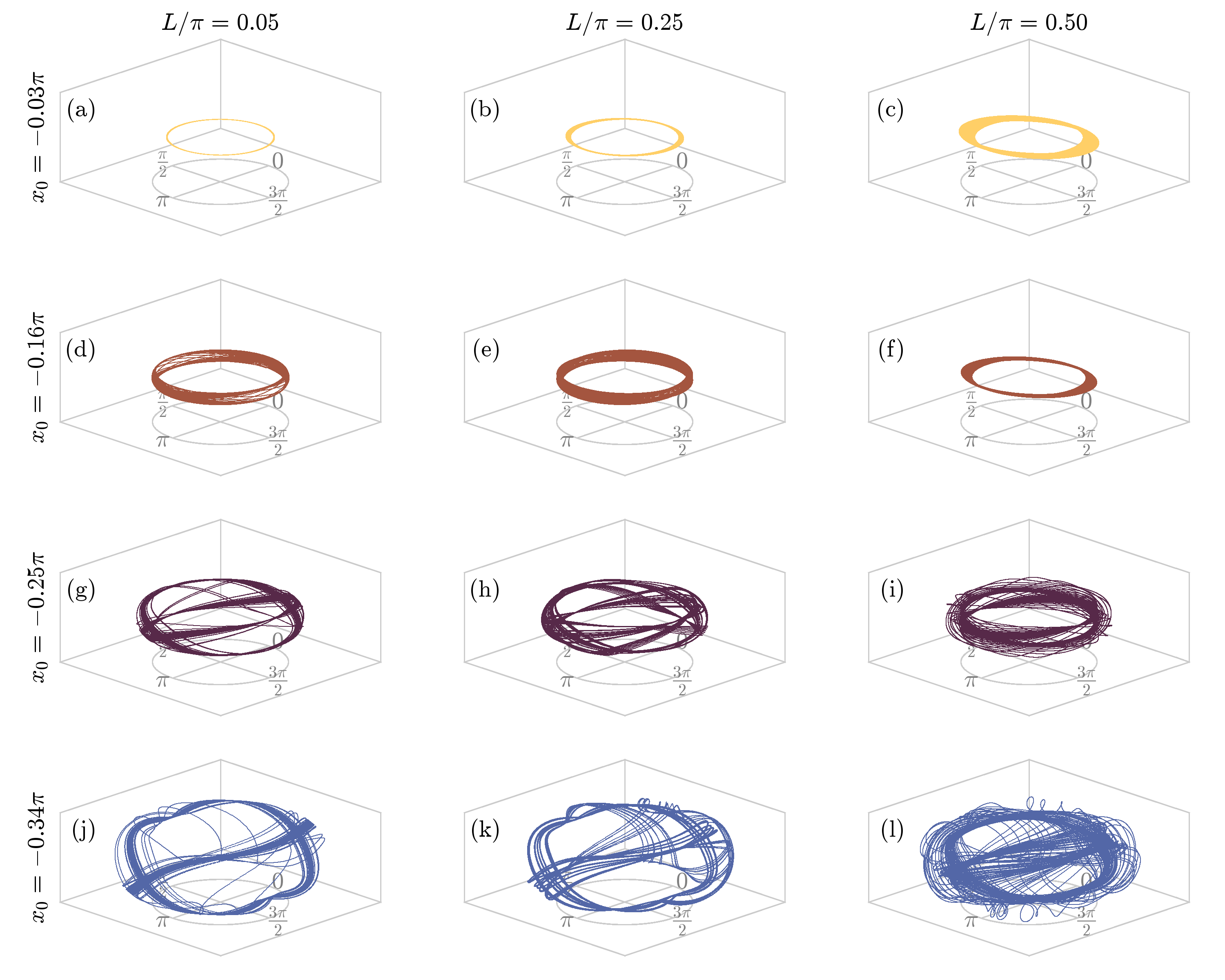}
    \caption{
    Trajectories corresponding to tori plotted in Fig.~\ref{fig:Poincare_3D}. Initial positions are as follows: $x_0$ varies by row, $y_0=0$, and $\phi_0=0$. The circular compass on the base of the plot denotes the angle $\phi$ around the tori. The radial position on each tori relates to the trajectory's position in the $x-y$ plane. 
    }
    \label{fig:Tori}
\end{figure}

We discuss in Subsection~\ref{sec:results:chaos} that some rod trajectories are quasiperiodic, which can be observed by the distinct tube-like shape their trajectories take in phase space (Fig.~\ref{fig:Poincare_3D}(d-f)). Because these trajectories are periodic in $\phi$, by connecting the ends of the tubes formed in phase space, we can form tori that the rod trajectories trace. We apply a change of variables to the $x$ and $y$ coordinates and plot the tori formed in Fig.~\ref{fig:Tori}. Each column represents a different rod length and each row a different initial condition. The rods that start near the center of the cell (yellow) form tight tori. The smallest rod's torus (Fig.~\ref{fig:Tori}(a)) is level, while the larger rods have tori which are skewed. The size of the torus increases with initial position. Additionally, the shape changes with the length of the rod, where larger rods trace a more irregular shape. 

\begin{figure}[tb]
    \centering
    \includegraphics[width=0.8\linewidth]{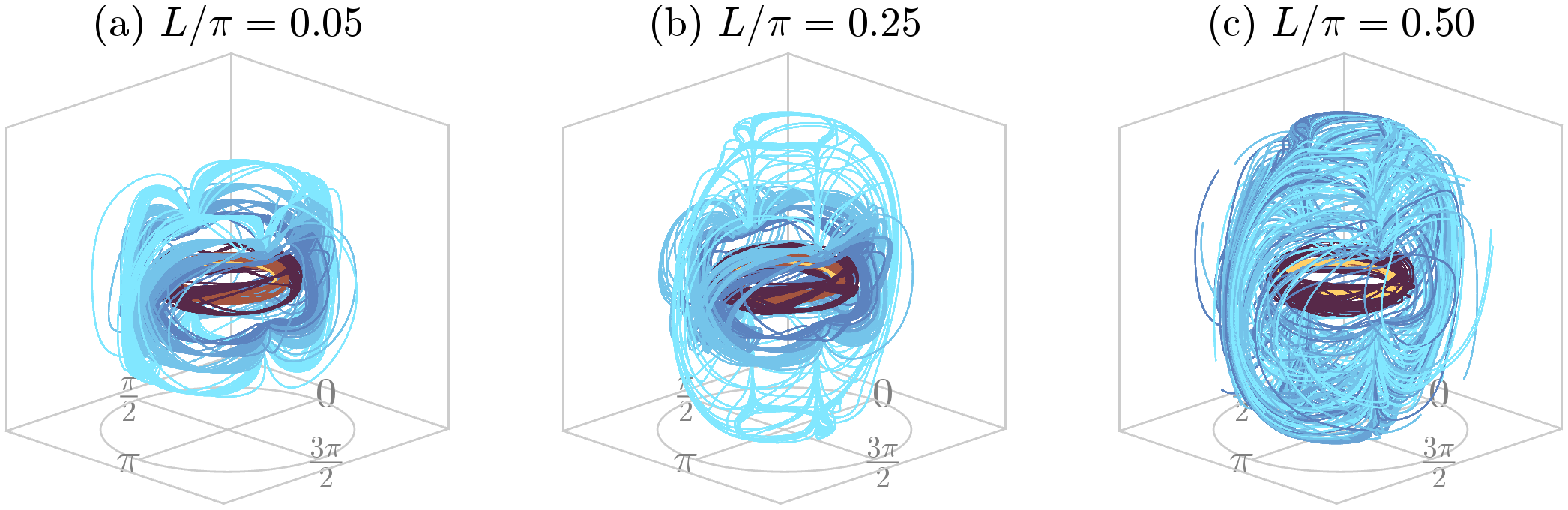}
    \caption{
    Chaotic trajectories from Fig.~\ref{fig:Poincare_3D} plotted using the change of variables used to form tori. Some quasi-periodic orbits are plotted for reference. 
    }
    \label{fig:Tori_chaos}
\end{figure}

These plots in Fig.~\ref{fig:Tori} do not display any chaotic trajectories. For reference, we show chaotic trajectories which have the same change of variables applied in Fig.~\ref{fig:Tori_chaos}. 

\bibliography{main} 

\begin{thebibliography}{45}%
\makeatletter
\providecommand \@ifxundefined [1]{%
 \@ifx{#1\undefined}
}%
\providecommand \@ifnum [1]{%
 \ifnum #1\expandafter \@firstoftwo
 \else \expandafter \@secondoftwo
 \fi
}%
\providecommand \@ifx [1]{%
 \ifx #1\expandafter \@firstoftwo
 \else \expandafter \@secondoftwo
 \fi
}%
\providecommand \natexlab [1]{#1}%
\providecommand \enquote  [1]{``#1''}%
\providecommand \bibnamefont  [1]{#1}%
\providecommand \bibfnamefont [1]{#1}%
\providecommand \citenamefont [1]{#1}%
\providecommand \href@noop [0]{\@secondoftwo}%
\providecommand \href [0]{\begingroup \@sanitize@url \@href}%
\providecommand \@href[1]{\@@startlink{#1}\@@href}%
\providecommand \@@href[1]{\endgroup#1\@@endlink}%
\providecommand \@sanitize@url [0]{\catcode `\\12\catcode `\$12\catcode `\&12\catcode `\#12\catcode `\^12\catcode `\_12\catcode `\%12\relax}%
\providecommand \@@startlink[1]{}%
\providecommand \@@endlink[0]{}%
\providecommand \url  [0]{\begingroup\@sanitize@url \@url }%
\providecommand \@url [1]{\endgroup\@href {#1}{\urlprefix }}%
\providecommand \urlprefix  [0]{URL }%
\providecommand \Eprint [0]{\href }%
\providecommand \doibase [0]{https://doi.org/}%
\providecommand \selectlanguage [0]{\@gobble}%
\providecommand \bibinfo  [0]{\@secondoftwo}%
\providecommand \bibfield  [0]{\@secondoftwo}%
\providecommand \translation [1]{[#1]}%
\providecommand \BibitemOpen [0]{}%
\providecommand \bibitemStop [0]{}%
\providecommand \bibitemNoStop [0]{.\EOS\space}%
\providecommand \EOS [0]{\spacefactor3000\relax}%
\providecommand \BibitemShut  [1]{\csname bibitem#1\endcsname}%
\let\auto@bib@innerbib\@empty
\bibitem [{\citenamefont {Harms}\ \emph {et~al.}(2024)\citenamefont {Harms}, \citenamefont {Brunton},\ and\ \citenamefont {McKeon}}]{harms_lagrangian_2024}%
  \BibitemOpen
  \bibfield  {author} {\bibinfo {author} {\bibfnamefont {T.~D.}\ \bibnamefont {Harms}}, \bibinfo {author} {\bibfnamefont {S.~L.}\ \bibnamefont {Brunton}},\ and\ \bibinfo {author} {\bibfnamefont {B.~J.}\ \bibnamefont {McKeon}},\ }\bibfield  {title} {\bibinfo {title} {Lagrangian gradient regression for the detection of coherent structures from sparse trajectory data},\ }\href {https://doi.org/10.1098/rsos.240586} {\bibfield  {journal} {\bibinfo  {journal} {R. Soc. Open Sci.}\ }\textbf {\bibinfo {volume} {11}},\ \bibinfo {pages} {240586} (\bibinfo {year} {2024})}\BibitemShut {NoStop}%
\bibitem [{\citenamefont {Manucharyan}\ \emph {et~al.}(2022)\citenamefont {Manucharyan}, \citenamefont {{Lopez-Acosta}},\ and\ \citenamefont {Wilhelmus}}]{manucharyan_spinning_2022}%
  \BibitemOpen
  \bibfield  {author} {\bibinfo {author} {\bibfnamefont {G.~E.}\ \bibnamefont {Manucharyan}}, \bibinfo {author} {\bibfnamefont {R.}~\bibnamefont {{Lopez-Acosta}}},\ and\ \bibinfo {author} {\bibfnamefont {M.~M.}\ \bibnamefont {Wilhelmus}},\ }\bibfield  {title} {\bibinfo {title} {Spinning ice floes reveal intensification of mesoscale eddies in the western {{Arctic Ocean}}},\ }\href {https://doi.org/10.1038/s41598-022-10712-z} {\bibfield  {journal} {\bibinfo  {journal} {Sci Rep}\ }\textbf {\bibinfo {volume} {12}},\ \bibinfo {pages} {7070} (\bibinfo {year} {2022})}\BibitemShut {NoStop}%
\bibitem [{\citenamefont {{Lopez-Acosta}}\ \emph {et~al.}(2019)\citenamefont {{Lopez-Acosta}}, \citenamefont {Schodlok},\ and\ \citenamefont {Wilhelmus}}]{lopez-acosta_ice_2019}%
  \BibitemOpen
  \bibfield  {author} {\bibinfo {author} {\bibfnamefont {R.}~\bibnamefont {{Lopez-Acosta}}}, \bibinfo {author} {\bibfnamefont {M.}~\bibnamefont {Schodlok}},\ and\ \bibinfo {author} {\bibfnamefont {M.}~\bibnamefont {Wilhelmus}},\ }\bibfield  {title} {\bibinfo {title} {Ice {{Floe Tracker}}: {{An}} algorithm to automatically retrieve {{Lagrangian}} trajectories via feature matching from moderate-resolution visual imagery},\ }\href {https://doi.org/10.1016/j.rse.2019.111406} {\bibfield  {journal} {\bibinfo  {journal} {Remote Sens. Environ.}\ }\textbf {\bibinfo {volume} {234}},\ \bibinfo {pages} {111406} (\bibinfo {year} {2019})}\BibitemShut {NoStop}%
\bibitem [{\citenamefont {Ghaffarian}\ \emph {et~al.}(2020)\citenamefont {Ghaffarian}, \citenamefont {Lopez}, \citenamefont {Mignot}, \citenamefont {Piegay},\ and\ \citenamefont {Riviere}}]{ghaffarian_dynamics_2020}%
  \BibitemOpen
  \bibfield  {author} {\bibinfo {author} {\bibfnamefont {H.}~\bibnamefont {Ghaffarian}}, \bibinfo {author} {\bibfnamefont {D.}~\bibnamefont {Lopez}}, \bibinfo {author} {\bibfnamefont {E.}~\bibnamefont {Mignot}}, \bibinfo {author} {\bibfnamefont {H.}~\bibnamefont {Piegay}},\ and\ \bibinfo {author} {\bibfnamefont {N.}~\bibnamefont {Riviere}},\ }\bibfield  {title} {\bibinfo {title} {Dynamics of floating objects at high particulate {{Reynolds}} numbers},\ }\href {https://doi.org/10.1103/PhysRevFluids.5.054307} {\bibfield  {journal} {\bibinfo  {journal} {Phys. Rev. Fluids}\ }\textbf {\bibinfo {volume} {5}},\ \bibinfo {pages} {054307} (\bibinfo {year} {2020})}\BibitemShut {NoStop}%
\bibitem [{\citenamefont {Qureshi}\ \emph {et~al.}(2007)\citenamefont {Qureshi}, \citenamefont {Bourgoin}, \citenamefont {Baudet}, \citenamefont {Cartellier},\ and\ \citenamefont {Gagne}}]{qureshi_turbulent_2007}%
  \BibitemOpen
  \bibfield  {author} {\bibinfo {author} {\bibfnamefont {N.~M.}\ \bibnamefont {Qureshi}}, \bibinfo {author} {\bibfnamefont {M.}~\bibnamefont {Bourgoin}}, \bibinfo {author} {\bibfnamefont {C.}~\bibnamefont {Baudet}}, \bibinfo {author} {\bibfnamefont {A.}~\bibnamefont {Cartellier}},\ and\ \bibinfo {author} {\bibfnamefont {Y.}~\bibnamefont {Gagne}},\ }\bibfield  {title} {\bibinfo {title} {Turbulent {{Transport}} of {{Material Particles}}: {{An Experimental Study}} of {{Finite Size Effects}}},\ }\href {https://doi.org/10.1103/PhysRevLett.99.184502} {\bibfield  {journal} {\bibinfo  {journal} {Phys. Rev. Lett.}\ }\textbf {\bibinfo {volume} {99}},\ \bibinfo {pages} {184502} (\bibinfo {year} {2007})}\BibitemShut {NoStop}%
\bibitem [{\citenamefont {Volk}\ \emph {et~al.}(2011)\citenamefont {Volk}, \citenamefont {Calzavarini}, \citenamefont {Leveque},\ and\ \citenamefont {Pinton}}]{volk_dynamics_2011}%
  \BibitemOpen
  \bibfield  {author} {\bibinfo {author} {\bibfnamefont {R.}~\bibnamefont {Volk}}, \bibinfo {author} {\bibfnamefont {E.}~\bibnamefont {Calzavarini}}, \bibinfo {author} {\bibfnamefont {E.}~\bibnamefont {Leveque}},\ and\ \bibinfo {author} {\bibfnamefont {J.-F.}\ \bibnamefont {Pinton}},\ }\bibfield  {title} {\bibinfo {title} {Dynamics of inertial particles in a turbulent von {{K{\'a}rm{\'a}n}} flow},\ }\href {https://doi.org/10.1017/S0022112010005690} {\bibfield  {journal} {\bibinfo  {journal} {J. Fluid Mech.}\ }\textbf {\bibinfo {volume} {668}},\ \bibinfo {pages} {223–235} (\bibinfo {year} {2011})}\BibitemShut {NoStop}%
\bibitem [{\citenamefont {Brandt}\ and\ \citenamefont {Coletti}(2022)}]{brandt_particleladen_2022}%
  \BibitemOpen
  \bibfield  {author} {\bibinfo {author} {\bibfnamefont {L.}~\bibnamefont {Brandt}}\ and\ \bibinfo {author} {\bibfnamefont {F.}~\bibnamefont {Coletti}},\ }\bibfield  {title} {\bibinfo {title} {Particle-{{Laden Turbulence}}: {{Progress}} and {{Perspectives}}},\ }\href {https://doi.org/10.1146/annurev-fluid-030121-021103} {\bibfield  {journal} {\bibinfo  {journal} {Annu. Rev. Fluid Mech.}\ }\textbf {\bibinfo {volume} {54}},\ \bibinfo {pages} {159} (\bibinfo {year} {2022})}\BibitemShut {NoStop}%
\bibitem [{\citenamefont {Fiabane}\ \emph {et~al.}(2012)\citenamefont {Fiabane}, \citenamefont {Zimmermann}, \citenamefont {Volk}, \citenamefont {Pinton},\ and\ \citenamefont {Bourgoin}}]{fiabane_clustering_2012}%
  \BibitemOpen
  \bibfield  {author} {\bibinfo {author} {\bibfnamefont {L.}~\bibnamefont {Fiabane}}, \bibinfo {author} {\bibfnamefont {R.}~\bibnamefont {Zimmermann}}, \bibinfo {author} {\bibfnamefont {R.}~\bibnamefont {Volk}}, \bibinfo {author} {\bibfnamefont {J.-F.}\ \bibnamefont {Pinton}},\ and\ \bibinfo {author} {\bibfnamefont {M.}~\bibnamefont {Bourgoin}},\ }\bibfield  {title} {\bibinfo {title} {Clustering of finite-size particles in turbulence},\ }\href {https://doi.org/10.1103/PhysRevE.86.035301} {\bibfield  {journal} {\bibinfo  {journal} {Phys. Rev. E}\ }\textbf {\bibinfo {volume} {86}},\ \bibinfo {pages} {035301} (\bibinfo {year} {2012})}\BibitemShut {NoStop}%
\bibitem [{\citenamefont {Lepp{\"a}ranta}\ \emph {et~al.}(2012)\citenamefont {Lepp{\"a}ranta}, \citenamefont {Oikkonen}, \citenamefont {Shirasawa},\ and\ \citenamefont {Fukamachi}}]{lepparanta_treatise_2012}%
  \BibitemOpen
  \bibfield  {author} {\bibinfo {author} {\bibfnamefont {M.}~\bibnamefont {Lepp{\"a}ranta}}, \bibinfo {author} {\bibfnamefont {A.}~\bibnamefont {Oikkonen}}, \bibinfo {author} {\bibfnamefont {K.}~\bibnamefont {Shirasawa}},\ and\ \bibinfo {author} {\bibfnamefont {Y.}~\bibnamefont {Fukamachi}},\ }\bibfield  {title} {\bibinfo {title} {A treatise on frequency spectrum of drift ice velocity},\ }\href {https://doi.org/10.1016/j.coldregions.2011.12.005} {\bibfield  {journal} {\bibinfo  {journal} {Cold Reg. Sci. Technol.}\ }\textbf {\bibinfo {volume} {76}},\ \bibinfo {pages} {83} (\bibinfo {year} {2012})}\BibitemShut {NoStop}%
\bibitem [{\citenamefont {Brenner}\ \emph {et~al.}(2023)\citenamefont {Brenner}, \citenamefont {Horvat}, \citenamefont {Hall}, \citenamefont {Lo~Piccolo}, \citenamefont {Fox-Kemper}, \citenamefont {Labb{\'e}},\ and\ \citenamefont {Dansereau}}]{brenner_scaledependent_2023}%
  \BibitemOpen
  \bibfield  {author} {\bibinfo {author} {\bibfnamefont {S.}~\bibnamefont {Brenner}}, \bibinfo {author} {\bibfnamefont {C.}~\bibnamefont {Horvat}}, \bibinfo {author} {\bibfnamefont {P.}~\bibnamefont {Hall}}, \bibinfo {author} {\bibfnamefont {A.}~\bibnamefont {Lo~Piccolo}}, \bibinfo {author} {\bibfnamefont {B.}~\bibnamefont {Fox-Kemper}}, \bibinfo {author} {\bibfnamefont {S.}~\bibnamefont {Labb{\'e}}},\ and\ \bibinfo {author} {\bibfnamefont {V.}~\bibnamefont {Dansereau}},\ }\bibfield  {title} {\bibinfo {title} {Scale-dependent air-sea exchange in the polar oceans: Floe-floe and floe-flow coupling in the generation of ice-ocean boundary layer turbulence},\ }\href {https://doi.org/10.1029/2023GL105703} {\bibfield  {journal} {\bibinfo  {journal} {Geophys. Res. Lett.}\ }\textbf {\bibinfo {volume} {50}},\ \bibinfo {pages} {e2023GL105703} (\bibinfo {year} {2023})}\BibitemShut {NoStop}%
\bibitem [{\citenamefont {Kim}\ \emph {et~al.}(2025)\citenamefont {Kim}, \citenamefont {Manucharyan},\ and\ \citenamefont {Wilhelmus}}]{kim_characterisation_2025}%
  \BibitemOpen
  \bibfield  {author} {\bibinfo {author} {\bibfnamefont {M.}~\bibnamefont {Kim}}, \bibinfo {author} {\bibfnamefont {G.~E.}\ \bibnamefont {Manucharyan}},\ and\ \bibinfo {author} {\bibfnamefont {M.~M.}\ \bibnamefont {Wilhelmus}},\ }\bibfield  {title} {\bibinfo {title} {Characterisation of sea ice kinematics over oceanic eddies},\ }\href {https://doi.org/10.1017/jfm.2025.10213} {\bibfield  {journal} {\bibinfo  {journal} {J. Fluid Mech.}\ }\textbf {\bibinfo {volume} {1015}},\ \bibinfo {pages} {A51} (\bibinfo {year} {2025})}\BibitemShut {NoStop}%
\bibitem [{\citenamefont {Jeffery}(1922)}]{jeffery_motion_1922}%
  \BibitemOpen
  \bibfield  {author} {\bibinfo {author} {\bibfnamefont {G.}~\bibnamefont {Jeffery}},\ }\bibfield  {title} {\bibinfo {title} {The motion of ellipsoidal particles immersed in a viscous fluid},\ }\href {https://doi.org/10.1098/rspa.1922.0078} {\bibfield  {journal} {\bibinfo  {journal} {Proc. R. Soc. Lond.}\ }\bibinfo {series} {A},\ \textbf {\bibinfo {volume} {102}},\ \bibinfo {pages} {161} (\bibinfo {year} {1922})}\BibitemShut {NoStop}%
\bibitem [{\citenamefont {Parsa}\ \emph {et~al.}(2012)\citenamefont {Parsa}, \citenamefont {Calzavarini}, \citenamefont {Toschi},\ and\ \citenamefont {Voth}}]{parsa_rotation_2012}%
  \BibitemOpen
  \bibfield  {author} {\bibinfo {author} {\bibfnamefont {S.}~\bibnamefont {Parsa}}, \bibinfo {author} {\bibfnamefont {E.}~\bibnamefont {Calzavarini}}, \bibinfo {author} {\bibfnamefont {F.}~\bibnamefont {Toschi}},\ and\ \bibinfo {author} {\bibfnamefont {G.~A.}\ \bibnamefont {Voth}},\ }\bibfield  {title} {\bibinfo {title} {Rotation {{Rate}} of {{Rods}} in {{Turbulent Fluid Flow}}},\ }\href {https://doi.org/10.1103/PhysRevLett.109.134501} {\bibfield  {journal} {\bibinfo  {journal} {Phys. Rev. Lett.}\ }\textbf {\bibinfo {volume} {109}},\ \bibinfo {pages} {134501} (\bibinfo {year} {2012})}\BibitemShut {NoStop}%
\bibitem [{\citenamefont {Pumir}\ and\ \citenamefont {Wilkinson}(2011)}]{pumir_orientation_2011}%
  \BibitemOpen
  \bibfield  {author} {\bibinfo {author} {\bibfnamefont {A.}~\bibnamefont {Pumir}}\ and\ \bibinfo {author} {\bibfnamefont {M.}~\bibnamefont {Wilkinson}},\ }\bibfield  {title} {\bibinfo {title} {Orientation statistics of small particles in turbulence},\ }\href {https://doi.org/10.1088/1367-2630/13/9/093030} {\bibfield  {journal} {\bibinfo  {journal} {New J. Phys.}\ }\textbf {\bibinfo {volume} {13}},\ \bibinfo {pages} {093030} (\bibinfo {year} {2011})}\BibitemShut {NoStop}%
\bibitem [{\citenamefont {Ni}\ \emph {et~al.}(2015)\citenamefont {Ni}, \citenamefont {Kramel}, \citenamefont {Ouellette},\ and\ \citenamefont {Voth}}]{ni_measurements_2015}%
  \BibitemOpen
  \bibfield  {author} {\bibinfo {author} {\bibfnamefont {R.}~\bibnamefont {Ni}}, \bibinfo {author} {\bibfnamefont {S.}~\bibnamefont {Kramel}}, \bibinfo {author} {\bibfnamefont {N.~T.}\ \bibnamefont {Ouellette}},\ and\ \bibinfo {author} {\bibfnamefont {G.~A.}\ \bibnamefont {Voth}},\ }\bibfield  {title} {\bibinfo {title} {Measurements of the coupling between the tumbling of rods and the velocity gradient tensor in turbulence},\ }\href {https://doi.org/10.1017/jfm.2015.16} {\bibfield  {journal} {\bibinfo  {journal} {J. Fluid Mech.}\ }\textbf {\bibinfo {volume} {766}},\ \bibinfo {pages} {202} (\bibinfo {year} {2015})}\BibitemShut {NoStop}%
\bibitem [{\citenamefont {Marcus}\ \emph {et~al.}(2014)\citenamefont {Marcus}, \citenamefont {Parsa}, \citenamefont {Kramel}, \citenamefont {Ni},\ and\ \citenamefont {Voth}}]{marcus_measurements_2014}%
  \BibitemOpen
  \bibfield  {author} {\bibinfo {author} {\bibfnamefont {G.~G.}\ \bibnamefont {Marcus}}, \bibinfo {author} {\bibfnamefont {S.}~\bibnamefont {Parsa}}, \bibinfo {author} {\bibfnamefont {S.}~\bibnamefont {Kramel}}, \bibinfo {author} {\bibfnamefont {R.}~\bibnamefont {Ni}},\ and\ \bibinfo {author} {\bibfnamefont {G.~A.}\ \bibnamefont {Voth}},\ }\bibfield  {title} {\bibinfo {title} {Measurements of the solid-body rotation of anisotropic particles in {{3D}} turbulence},\ }\href {https://doi.org/10.1088/1367-2630/16/10/102001} {\bibfield  {journal} {\bibinfo  {journal} {New J. Phys.}\ }\textbf {\bibinfo {volume} {16}},\ \bibinfo {pages} {102001} (\bibinfo {year} {2014})}\BibitemShut {NoStop}%
\bibitem [{\citenamefont {Cavaiola}\ \emph {et~al.}(2020)\citenamefont {Cavaiola}, \citenamefont {Olivieri},\ and\ \citenamefont {Mazzino}}]{cavaiola_assembly_2020}%
  \BibitemOpen
  \bibfield  {author} {\bibinfo {author} {\bibfnamefont {M.}~\bibnamefont {Cavaiola}}, \bibinfo {author} {\bibfnamefont {S.}~\bibnamefont {Olivieri}},\ and\ \bibinfo {author} {\bibfnamefont {A.}~\bibnamefont {Mazzino}},\ }\bibfield  {title} {\bibinfo {title} {The assembly of freely moving rigid fibres measures the flow velocity gradient tensor},\ }\href {https://doi.org/10.1017/jfm.2020.288} {\bibfield  {journal} {\bibinfo  {journal} {J. Fluid Mech.}\ }\textbf {\bibinfo {volume} {894}},\ \bibinfo {pages} {A25} (\bibinfo {year} {2020})}\BibitemShut {NoStop}%
\bibitem [{\citenamefont {Brizzolara}\ \emph {et~al.}(2021)\citenamefont {Brizzolara}, \citenamefont {Rosti}, \citenamefont {Olivieri}, \citenamefont {Brandt}, \citenamefont {Holzner},\ and\ \citenamefont {Mazzino}}]{brizzolara_fiber_2021}%
  \BibitemOpen
  \bibfield  {author} {\bibinfo {author} {\bibfnamefont {S.}~\bibnamefont {Brizzolara}}, \bibinfo {author} {\bibfnamefont {M.~E.}\ \bibnamefont {Rosti}}, \bibinfo {author} {\bibfnamefont {S.}~\bibnamefont {Olivieri}}, \bibinfo {author} {\bibfnamefont {L.}~\bibnamefont {Brandt}}, \bibinfo {author} {\bibfnamefont {M.}~\bibnamefont {Holzner}},\ and\ \bibinfo {author} {\bibfnamefont {A.}~\bibnamefont {Mazzino}},\ }\bibfield  {title} {\bibinfo {title} {Fiber {{Tracking Velocimetry}} for {{Two-Point Statistics}} of {{Turbulence}}},\ }\href {https://doi.org/10.1103/PhysRevX.11.031060} {\bibfield  {journal} {\bibinfo  {journal} {Phys. Rev. X}\ }\textbf {\bibinfo {volume} {11}},\ \bibinfo {pages} {031060} (\bibinfo {year} {2021})}\BibitemShut {NoStop}%
\bibitem [{\citenamefont {Sudharsan}\ \emph {et~al.}(2016)\citenamefont {Sudharsan}, \citenamefont {Brunton},\ and\ \citenamefont {Riley}}]{sudharsan_lagrangian_2016}%
  \BibitemOpen
  \bibfield  {author} {\bibinfo {author} {\bibfnamefont {M.}~\bibnamefont {Sudharsan}}, \bibinfo {author} {\bibfnamefont {S.~L.}\ \bibnamefont {Brunton}},\ and\ \bibinfo {author} {\bibfnamefont {J.~J.}\ \bibnamefont {Riley}},\ }\bibfield  {title} {\bibinfo {title} {Lagrangian coherent structures and inertial particle dynamics},\ }\href {https://doi.org/10.1103/PhysRevE.93.033108} {\bibfield  {journal} {\bibinfo  {journal} {Phys. Rev. E}\ }\textbf {\bibinfo {volume} {93}},\ \bibinfo {pages} {033108} (\bibinfo {year} {2016})}\BibitemShut {NoStop}%
\bibitem [{\citenamefont {Haller}\ and\ \citenamefont {Sapsis}(2008)}]{haller_inertial_2008}%
  \BibitemOpen
  \bibfield  {author} {\bibinfo {author} {\bibfnamefont {G.}~\bibnamefont {Haller}}\ and\ \bibinfo {author} {\bibfnamefont {T.}~\bibnamefont {Sapsis}},\ }\bibfield  {title} {\bibinfo {title} {Where do inertial particles go in fluid flows?},\ }\href {https://doi.org/10.1016/j.physd.2007.09.027} {\bibfield  {journal} {\bibinfo  {journal} {Phys. D: Nonlinear Phenom.}\ }\textbf {\bibinfo {volume} {237}},\ \bibinfo {pages} {573} (\bibinfo {year} {2008})}\BibitemShut {NoStop}%
\bibitem [{\citenamefont {Sapsis}\ and\ \citenamefont {Haller}(2009)}]{sapsis_inertial_2009}%
  \BibitemOpen
  \bibfield  {author} {\bibinfo {author} {\bibfnamefont {T.}~\bibnamefont {Sapsis}}\ and\ \bibinfo {author} {\bibfnamefont {G.}~\bibnamefont {Haller}},\ }\bibfield  {title} {\bibinfo {title} {Inertial particle dynamics in a hurricane},\ }\href {https://doi.org/10.1175/2009JAS2865.1} {\bibfield  {journal} {\bibinfo  {journal} {J. Atmos. Sci.}\ }\textbf {\bibinfo {volume} {66}},\ \bibinfo {pages} {2481–2492} (\bibinfo {year} {2009})}\BibitemShut {NoStop}%
\bibitem [{\citenamefont {Eaton}\ and\ \citenamefont {Fessler}(1994)}]{eaton_preferential_1994}%
  \BibitemOpen
  \bibfield  {author} {\bibinfo {author} {\bibfnamefont {J.}~\bibnamefont {Eaton}}\ and\ \bibinfo {author} {\bibfnamefont {J.}~\bibnamefont {Fessler}},\ }\bibfield  {title} {\bibinfo {title} {Preferential concentration of particles by turbulence},\ }\href {https://doi.org/10.1016/0301-9322(94)90072-8} {\bibfield  {journal} {\bibinfo  {journal} {Int. J. Multiph. Flow}\ }\textbf {\bibinfo {volume} {20}},\ \bibinfo {pages} {169} (\bibinfo {year} {1994})}\BibitemShut {NoStop}%
\bibitem [{\citenamefont {Squires}\ and\ \citenamefont {Eaton}(1991)}]{squires_measurements_1991}%
  \BibitemOpen
  \bibfield  {author} {\bibinfo {author} {\bibfnamefont {K.~D.}\ \bibnamefont {Squires}}\ and\ \bibinfo {author} {\bibfnamefont {J.~K.}\ \bibnamefont {Eaton}},\ }\bibfield  {title} {\bibinfo {title} {Measurements of particle dispersion obtained from direct numerical simulations of isotropic turbulence},\ }\href {https://doi.org/10.1017/S0022112091002276} {\bibfield  {journal} {\bibinfo  {journal} {J. Fluid Mech.}\ }\textbf {\bibinfo {volume} {226}},\ \bibinfo {pages} {1} (\bibinfo {year} {1991})}\BibitemShut {NoStop}%
\bibitem [{\citenamefont {Mathai}\ \emph {et~al.}(2020)\citenamefont {Mathai}, \citenamefont {Lohse},\ and\ \citenamefont {Sun}}]{mathai_bubbly_2020}%
  \BibitemOpen
  \bibfield  {author} {\bibinfo {author} {\bibfnamefont {V.}~\bibnamefont {Mathai}}, \bibinfo {author} {\bibfnamefont {D.}~\bibnamefont {Lohse}},\ and\ \bibinfo {author} {\bibfnamefont {C.}~\bibnamefont {Sun}},\ }\bibfield  {title} {\bibinfo {title} {Bubbly and {{Buoyant Particle}}--{{Laden Turbulent Flows}}},\ }\href {https://doi.org/10.1146/annurev-conmatphys-031119-050637} {\bibfield  {journal} {\bibinfo  {journal} {Annu. Rev. Condens. Matter Phys.}\ }\textbf {\bibinfo {volume} {11}},\ \bibinfo {pages} {529} (\bibinfo {year} {2020})}\BibitemShut {NoStop}%
\bibitem [{\citenamefont {Zhang}\ \emph {et~al.}(2001)\citenamefont {Zhang}, \citenamefont {Ahmadi}, \citenamefont {Fan},\ and\ \citenamefont {McLaughlin}}]{zhang_ellipsoidal_2001}%
  \BibitemOpen
  \bibfield  {author} {\bibinfo {author} {\bibfnamefont {H.}~\bibnamefont {Zhang}}, \bibinfo {author} {\bibfnamefont {G.}~\bibnamefont {Ahmadi}}, \bibinfo {author} {\bibfnamefont {F.-G.}\ \bibnamefont {Fan}},\ and\ \bibinfo {author} {\bibfnamefont {J.~B.}\ \bibnamefont {McLaughlin}},\ }\bibfield  {title} {\bibinfo {title} {Ellipsoidal particles transport and deposition in turbulent channel fows},\ }\href {https://doi.org/https://doi.org/10.1016/S0301-9322(00)00064-1} {\bibfield  {journal} {\bibinfo  {journal} {Int. J. Multiph. Flow}\ }\textbf {\bibinfo {volume} {27}},\ \bibinfo {pages} {971} (\bibinfo {year} {2001})}\BibitemShut {NoStop}%
\bibitem [{\citenamefont {Mortensen}\ \emph {et~al.}(2008)\citenamefont {Mortensen}, \citenamefont {Andersson}, \citenamefont {Gillissen},\ and\ \citenamefont {Boersma}}]{mortensen_dynamics_2008}%
  \BibitemOpen
  \bibfield  {author} {\bibinfo {author} {\bibfnamefont {P.~H.}\ \bibnamefont {Mortensen}}, \bibinfo {author} {\bibfnamefont {H.~I.}\ \bibnamefont {Andersson}}, \bibinfo {author} {\bibfnamefont {J.~J.~J.}\ \bibnamefont {Gillissen}},\ and\ \bibinfo {author} {\bibfnamefont {B.~J.}\ \bibnamefont {Boersma}},\ }\bibfield  {title} {\bibinfo {title} {Dynamics of prolate ellipsoidal particles in a turbulent channel flow},\ }\href {https://doi.org/10.1063/1.2975209} {\bibfield  {journal} {\bibinfo  {journal} {Phys. Fluids}\ }\textbf {\bibinfo {volume} {20}},\ \bibinfo {pages} {093302} (\bibinfo {year} {2008})}\BibitemShut {NoStop}%
\bibitem [{\citenamefont {Fujiki}\ \emph {et~al.}(2024)\citenamefont {Fujiki}, \citenamefont {Awai}, \citenamefont {Motoori},\ and\ \citenamefont {Goto}}]{fujiki_attraction_2024}%
  \BibitemOpen
  \bibfield  {author} {\bibinfo {author} {\bibfnamefont {Y.}~\bibnamefont {Fujiki}}, \bibinfo {author} {\bibfnamefont {H.}~\bibnamefont {Awai}}, \bibinfo {author} {\bibfnamefont {Y.}~\bibnamefont {Motoori}},\ and\ \bibinfo {author} {\bibfnamefont {S.}~\bibnamefont {Goto}},\ }\bibfield  {title} {\bibinfo {title} {Attraction of neutrally buoyant deformable particles towards a vortex},\ }\href {https://doi.org/10.1103/PhysRevFluids.9.014301} {\bibfield  {journal} {\bibinfo  {journal} {Phys. Rev. Fluids}\ }\textbf {\bibinfo {volume} {9}},\ \bibinfo {pages} {014301} (\bibinfo {year} {2024})}\BibitemShut {NoStop}%
\bibitem [{\citenamefont {Picardo}\ \emph {et~al.}(2020)\citenamefont {Picardo}, \citenamefont {Singh}, \citenamefont {Ray},\ and\ \citenamefont {Vincenzi}}]{picardo_dynamics_2020}%
  \BibitemOpen
  \bibfield  {author} {\bibinfo {author} {\bibfnamefont {J.~R.}\ \bibnamefont {Picardo}}, \bibinfo {author} {\bibfnamefont {R.}~\bibnamefont {Singh}}, \bibinfo {author} {\bibfnamefont {S.~S.}\ \bibnamefont {Ray}},\ and\ \bibinfo {author} {\bibfnamefont {D.}~\bibnamefont {Vincenzi}},\ }\bibfield  {title} {\bibinfo {title} {Dynamics of a long chain in turbulent flows: Impact of vortices},\ }\href {https://doi.org/10.1098/rsta.2019.0405} {\bibfield  {journal} {\bibinfo  {journal} {Phil. Trans. R. Soc. A.}\ }\textbf {\bibinfo {volume} {378}},\ \bibinfo {pages} {20190405} (\bibinfo {year} {2020})}\BibitemShut {NoStop}%
\bibitem [{\citenamefont {Atis}\ \emph {et~al.}(2022)\citenamefont {Atis}, \citenamefont {Leclair}, \citenamefont {Sapsis},\ and\ \citenamefont {Peacock}}]{atis_anisotropic_2022}%
  \BibitemOpen
  \bibfield  {author} {\bibinfo {author} {\bibfnamefont {S.}~\bibnamefont {Atis}}, \bibinfo {author} {\bibfnamefont {M.}~\bibnamefont {Leclair}}, \bibinfo {author} {\bibfnamefont {T.~P.}\ \bibnamefont {Sapsis}},\ and\ \bibinfo {author} {\bibfnamefont {T.}~\bibnamefont {Peacock}},\ }\bibfield  {title} {\bibinfo {title} {Anisotropic particles focusing effect in complex flows},\ }\href {https://doi.org/10.1103/PhysRevFluids.7.084503} {\bibfield  {journal} {\bibinfo  {journal} {Phys. Rev. Fluids}\ }\textbf {\bibinfo {volume} {7}},\ \bibinfo {pages} {084503} (\bibinfo {year} {2022})}\BibitemShut {NoStop}%
\bibitem [{\citenamefont {Lopez}\ and\ \citenamefont {Guazzelli}(2017)}]{lopez_inertial_2017}%
  \BibitemOpen
  \bibfield  {author} {\bibinfo {author} {\bibfnamefont {D.}~\bibnamefont {Lopez}}\ and\ \bibinfo {author} {\bibfnamefont {E.}~\bibnamefont {Guazzelli}},\ }\bibfield  {title} {\bibinfo {title} {Inertial effects on fibers settling in a vortical flow},\ }\href {https://doi.org/10.1103/PhysRevFluids.2.024306} {\bibfield  {journal} {\bibinfo  {journal} {Phys. Rev. Fluids}\ }\textbf {\bibinfo {volume} {2}},\ \bibinfo {pages} {024306} (\bibinfo {year} {2017})}\BibitemShut {NoStop}%
\bibitem [{\citenamefont {Batchelor}(1970)}]{batchelor_slenderbody_1970}%
  \BibitemOpen
  \bibfield  {author} {\bibinfo {author} {\bibfnamefont {G.~K.}\ \bibnamefont {Batchelor}},\ }\bibfield  {title} {\bibinfo {title} {Slender-body theory for particles of arbitrary cross-section in {{Stokes}} flow},\ }\href {https://doi.org/10.1017/S002211207000191X} {\bibfield  {journal} {\bibinfo  {journal} {J. Fluid Mech.}\ }\textbf {\bibinfo {volume} {44}},\ \bibinfo {pages} {419} (\bibinfo {year} {1970})}\BibitemShut {NoStop}%
\bibitem [{\citenamefont {Cox}(1970)}]{cox_motion_1970}%
  \BibitemOpen
  \bibfield  {author} {\bibinfo {author} {\bibfnamefont {R.~G.}\ \bibnamefont {Cox}},\ }\bibfield  {title} {\bibinfo {title} {The motion of long slender bodies in a viscous fluid {{Part}} 1. {{General}} theory},\ }\href {https://doi.org/10.1017/S002211207000215X} {\bibfield  {journal} {\bibinfo  {journal} {J. Fluid Mech.}\ }\textbf {\bibinfo {volume} {44}},\ \bibinfo {pages} {791} (\bibinfo {year} {1970})}\BibitemShut {NoStop}%
\bibitem [{\citenamefont {Olson}\ and\ \citenamefont {Kerekes}(1998)}]{olson_motion_1998}%
  \BibitemOpen
  \bibfield  {author} {\bibinfo {author} {\bibfnamefont {J.~A.}\ \bibnamefont {Olson}}\ and\ \bibinfo {author} {\bibfnamefont {R.~J.}\ \bibnamefont {Kerekes}},\ }\bibfield  {title} {\bibinfo {title} {The motion of fibres in turbulent flow},\ }\href {https://doi.org/10.1017/S0022112098002973} {\bibfield  {journal} {\bibinfo  {journal} {J. Fluid Mech.}\ }\textbf {\bibinfo {volume} {377}},\ \bibinfo {pages} {47} (\bibinfo {year} {1998})}\BibitemShut {NoStop}%
\bibitem [{\citenamefont {Shin}\ and\ \citenamefont {Koch}(2005)}]{shin_rotational_2005}%
  \BibitemOpen
  \bibfield  {author} {\bibinfo {author} {\bibfnamefont {M.}~\bibnamefont {Shin}}\ and\ \bibinfo {author} {\bibfnamefont {D.~L.}\ \bibnamefont {Koch}},\ }\bibfield  {title} {\bibinfo {title} {Rotational and translational dispersion of fibres in isotropic turbulent flows},\ }\href {https://doi.org/10.1017/S0022112005005690} {\bibfield  {journal} {\bibinfo  {journal} {J. Fluid Mech.}\ }\textbf {\bibinfo {volume} {540}},\ \bibinfo {pages} {143} (\bibinfo {year} {2005})}\BibitemShut {NoStop}%
\bibitem [{\citenamefont {Pujara}\ \emph {et~al.}(2019)\citenamefont {Pujara}, \citenamefont {Voth},\ and\ \citenamefont {Variano}}]{pujara_scaledependent_2019}%
  \BibitemOpen
  \bibfield  {author} {\bibinfo {author} {\bibfnamefont {N.}~\bibnamefont {Pujara}}, \bibinfo {author} {\bibfnamefont {G.~A.}\ \bibnamefont {Voth}},\ and\ \bibinfo {author} {\bibfnamefont {E.~A.}\ \bibnamefont {Variano}},\ }\bibfield  {title} {\bibinfo {title} {Scale-dependent alignment, tumbling and stretching of slender rods in isotropic turbulence},\ }\href {https://doi.org/10.1017/jfm.2018.866} {\bibfield  {journal} {\bibinfo  {journal} {J. Fluid Mech.}\ }\textbf {\bibinfo {volume} {860}},\ \bibinfo {pages} {465} (\bibinfo {year} {2019})}\BibitemShut {NoStop}%
\bibitem [{\citenamefont {Subramanian}\ and\ \citenamefont {Koch}(2005)}]{subramanian_inertial_2005}%
  \BibitemOpen
  \bibfield  {author} {\bibinfo {author} {\bibfnamefont {G.}~\bibnamefont {Subramanian}}\ and\ \bibinfo {author} {\bibfnamefont {D.~L.}\ \bibnamefont {Koch}},\ }\bibfield  {title} {\bibinfo {title} {Inertial effects on fibre motion in simple shear flow},\ }\href {https://doi.org/10.1017/s0022112005004829} {\bibfield  {journal} {\bibinfo  {journal} {J. Fluid Mech.}\ }\textbf {\bibinfo {volume} {535}},\ \bibinfo {pages} {383} (\bibinfo {year} {2005})}\BibitemShut {NoStop}%
\bibitem [{\citenamefont {Butler}\ and\ \citenamefont {Shaqfeh}(2002)}]{butler_dynamic_2002}%
  \BibitemOpen
  \bibfield  {author} {\bibinfo {author} {\bibfnamefont {J.~E.}\ \bibnamefont {Butler}}\ and\ \bibinfo {author} {\bibfnamefont {E.~S.~G.}\ \bibnamefont {Shaqfeh}},\ }\bibfield  {title} {\bibinfo {title} {Dynamic simulations of the inhomogeneous sedimentation of rigid fibres},\ }\href {https://doi.org/10.1017/s0022112002001544} {\bibfield  {journal} {\bibinfo  {journal} {J. Fluid Mech.}\ }\textbf {\bibinfo {volume} {468}},\ \bibinfo {pages} {205} (\bibinfo {year} {2002})}\BibitemShut {NoStop}%
\bibitem [{\citenamefont {Mackaplow}\ and\ \citenamefont {Shaqfeh}(1998)}]{mackaplow_numerical_1998}%
  \BibitemOpen
  \bibfield  {author} {\bibinfo {author} {\bibfnamefont {M.~B.}\ \bibnamefont {Mackaplow}}\ and\ \bibinfo {author} {\bibfnamefont {E.~S.~G.}\ \bibnamefont {Shaqfeh}},\ }\bibfield  {title} {\bibinfo {title} {A numerical study of the sedimentation of fibre suspensions},\ }\href {https://doi.org/10.1017/s0022112098002663} {\bibfield  {journal} {\bibinfo  {journal} {J. Fluid Mech.}\ }\textbf {\bibinfo {volume} {376}},\ \bibinfo {pages} {149} (\bibinfo {year} {1998})}\BibitemShut {NoStop}%
\bibitem [{\citenamefont {Hu}\ \emph {et~al.}()\citenamefont {Hu}, \citenamefont {Yang}, \citenamefont {Luo}, \citenamefont {Zhang},\ and\ \citenamefont {Man}}]{hu_anomalous_2025}%
  \BibitemOpen
  \bibfield  {author} {\bibinfo {author} {\bibfnamefont {S.}~\bibnamefont {Hu}}, \bibinfo {author} {\bibfnamefont {X.}~\bibnamefont {Yang}}, \bibinfo {author} {\bibfnamefont {N.}~\bibnamefont {Luo}}, \bibinfo {author} {\bibfnamefont {J.}~\bibnamefont {Zhang}},\ and\ \bibinfo {author} {\bibfnamefont {X.}~\bibnamefont {Man}},\ }\bibfield  {title} {\bibinfo {title} {Anomalous transport of elongated particles in oscillatory vortical flows}\ }\href {https://doi.org/arXiv:2508.09677} {arXiv:2508.09677}\BibitemShut {NoStop}%
\bibitem [{\citenamefont {Hu}\ \emph {et~al.}(2021)\citenamefont {Hu}, \citenamefont {Chu}, \citenamefont {Shelley},\ and\ \citenamefont {Zhang}}]{hu_levy_2021}%
  \BibitemOpen
  \bibfield  {author} {\bibinfo {author} {\bibfnamefont {S.-Y.}\ \bibnamefont {Hu}}, \bibinfo {author} {\bibfnamefont {J.-J.}\ \bibnamefont {Chu}}, \bibinfo {author} {\bibfnamefont {M.~J.}\ \bibnamefont {Shelley}},\ and\ \bibinfo {author} {\bibfnamefont {J.}~\bibnamefont {Zhang}},\ }\bibfield  {title} {\bibinfo {title} {L\'evy walks and path chaos in the dispersal of elongated structures moving across cellular vortical flows},\ }\href {https://doi.org/10.1103/PhysRevLett.127.074503} {\bibfield  {journal} {\bibinfo  {journal} {Phys. Rev. Lett.}\ }\textbf {\bibinfo {volume} {127}},\ \bibinfo {pages} {074503} (\bibinfo {year} {2021})}\BibitemShut {NoStop}%
\bibitem [{\citenamefont {Manikantan}\ and\ \citenamefont {Saintillan}(2013)}]{manikantan_subdiffusive_2013}%
  \BibitemOpen
  \bibfield  {author} {\bibinfo {author} {\bibfnamefont {H.}~\bibnamefont {Manikantan}}\ and\ \bibinfo {author} {\bibfnamefont {D.}~\bibnamefont {Saintillan}},\ }\bibfield  {title} {\bibinfo {title} {Subdiffusive transport of fluctuating elastic filaments in cellular flows},\ }\href {https://doi.org/doi.org/10.1063/1.4812794} {\bibfield  {journal} {\bibinfo  {journal} {Phys. Fluids}\ }\textbf {\bibinfo {volume} {25}},\ \bibinfo {pages} {073603} (\bibinfo {year} {2013})}\BibitemShut {NoStop}%
\bibitem [{\citenamefont {Crameri}(2018)}]{crameri_scientific_2018}%
  \BibitemOpen
  \bibfield  {author} {\bibinfo {author} {\bibfnamefont {F.}~\bibnamefont {Crameri}},\ }\href {https://doi.org/10.5281/zenodo.1243862} {\bibinfo {title} {Scientific colour maps}} (\bibinfo {year} {2018})\BibitemShut {NoStop}%
\bibitem [{sup()}]{supplemental}%
  \BibitemOpen
  \href@noop {} {}\bibinfo {note} {See Supplemental Material at [link tbd] for animations of rod trajectories.}\BibitemShut {Stop}%
\bibitem [{\citenamefont {Aref}\ \emph {et~al.}(2017)\citenamefont {Aref}, \citenamefont {Blake}, \citenamefont {Budi\ifmmode \check{s}\else \v{s}\fi{}i\ifmmode~\acute{c}\else \'{c}\fi{}}, \citenamefont {Cardoso}, \citenamefont {Cartwright}, \citenamefont {Clercx}, \citenamefont {El~Omari}, \citenamefont {Feudel}, \citenamefont {Golestanian}, \citenamefont {Gouillart}, \citenamefont {van Heijst}, \citenamefont {Krasnopolskaya}, \citenamefont {Le~Guer}, \citenamefont {MacKay}, \citenamefont {Meleshko}, \citenamefont {Metcalfe}, \citenamefont {Mezi\ifmmode~\acute{c}\else \'{c}\fi{}}, \citenamefont {de~Moura}, \citenamefont {Piro}, \citenamefont {Speetjens}, \citenamefont {Sturman}, \citenamefont {Thiffeault},\ and\ \citenamefont {Tuval}}]{aref_frontiers_2017}%
  \BibitemOpen
  \bibfield  {author} {\bibinfo {author} {\bibfnamefont {H.}~\bibnamefont {Aref}}, \bibinfo {author} {\bibfnamefont {J.~R.}\ \bibnamefont {Blake}}, \bibinfo {author} {\bibfnamefont {M.}~\bibnamefont {Budi\ifmmode \check{s}\else \v{s}\fi{}i\ifmmode~\acute{c}\else \'{c}\fi{}}}, \bibinfo {author} {\bibfnamefont {S.~S.~S.}\ \bibnamefont {Cardoso}}, \bibinfo {author} {\bibfnamefont {J.~H.~E.}\ \bibnamefont {Cartwright}}, \bibinfo {author} {\bibfnamefont {H.~J.~H.}\ \bibnamefont {Clercx}}, \bibinfo {author} {\bibfnamefont {K.}~\bibnamefont {El~Omari}}, \bibinfo {author} {\bibfnamefont {U.}~\bibnamefont {Feudel}}, \bibinfo {author} {\bibfnamefont {R.}~\bibnamefont {Golestanian}}, \bibinfo {author} {\bibfnamefont {E.}~\bibnamefont {Gouillart}}, \bibinfo {author} {\bibfnamefont {G.~F.}\ \bibnamefont {van Heijst}}, \bibinfo {author} {\bibfnamefont {T.~S.}\ \bibnamefont {Krasnopolskaya}}, \bibinfo {author} {\bibfnamefont {Y.}~\bibnamefont {Le~Guer}}, \bibinfo {author} {\bibfnamefont {R.~S.}\ \bibnamefont {MacKay}},
  \bibinfo {author} {\bibfnamefont {V.~V.}\ \bibnamefont {Meleshko}}, \bibinfo {author} {\bibfnamefont {G.}~\bibnamefont {Metcalfe}}, \bibinfo {author} {\bibfnamefont {I.}~\bibnamefont {Mezi\ifmmode~\acute{c}\else \'{c}\fi{}}}, \bibinfo {author} {\bibfnamefont {A.~P.~S.}\ \bibnamefont {de~Moura}}, \bibinfo {author} {\bibfnamefont {O.}~\bibnamefont {Piro}}, \bibinfo {author} {\bibfnamefont {M.~F.~M.}\ \bibnamefont {Speetjens}}, \bibinfo {author} {\bibfnamefont {R.}~\bibnamefont {Sturman}}, \bibinfo {author} {\bibfnamefont {J.-L.}\ \bibnamefont {Thiffeault}},\ and\ \bibinfo {author} {\bibfnamefont {I.}~\bibnamefont {Tuval}},\ }\bibfield  {title} {\bibinfo {title} {Frontiers of chaotic advection},\ }\href {https://doi.org/10.1103/RevModPhys.89.025007} {\bibfield  {journal} {\bibinfo  {journal} {Rev. Mod. Phys.}\ }\textbf {\bibinfo {volume} {89}},\ \bibinfo {pages} {025007} (\bibinfo {year} {2017})}\BibitemShut {NoStop}%
\bibitem [{\citenamefont {Ott}(2008)}]{ott_chaos_2008}%
  \BibitemOpen
  \bibfield  {author} {\bibinfo {author} {\bibfnamefont {E.}~\bibnamefont {Ott}},\ }\href@noop {} {\emph {\bibinfo {title} {Chaos in Dynamical Systems}}},\ \bibinfo {edition} {2nd}\ ed.\ (\bibinfo  {publisher} {Cambridge Univ. Press},\ \bibinfo {address} {Cambridge},\ \bibinfo {year} {2008})\BibitemShut {NoStop}%
\end{thebibliography}%

\end{document}